\documentclass{svjour3}                     
\smartqed  
\usepackage{amssymb}
\usepackage{graphicx}
\usepackage{natbib}
\bibpunct{(}{)}{;}{a}{}{,}

\usepackage{mathptmx}      
\journalname{SSRv}

\newcommand{\msun}{\,h^{-1}{\rm M}_\odot}

\newcommand{\mincir}{\raise
  -2.truept\hbox{\rlap{\hbox{$\sim$}}\raise5.truept \hbox{$<$}\ }}
\usepackage{amsmath,amssymb}   

\newcommand{\siml}{\raise
  -2.truept\hbox{\rlap{\hbox{$\sim$}}\raise5.truept \hbox{$<$}\ }}
\newcommand{\simg}{\raise
  -2.truept\hbox{\rlap{\hbox{$\sim$}}\raise5.truept \hbox{$>$}\ }}

\newcommand{\be}{\begin{equation}}
\newcommand{\ee}{\end{equation}}
\newcommand{\beq}{\begin{eqnarray}}
\newcommand{\eeq}{\end{eqnarray}}

\begin{document}

\title{The chemical enrichment of the ICM from hydrodynamical
  simulations}

\author{S.~Borgani \and D.~Fabjan \and L.~Tornatore \and S.~Schindler \and K.~Dolag \and A.~Diaferio}

\authorrunning{Borgani et al.}
\titlerunning{Simulations of the chemical enrichment of the ICM}

\institute{
S. Borgani \at
Department of Astronomy, University of Trieste, via Tiepolo 11, I-34143 Trieste, Italy\\
\email{borgani@oats.inaf.it}\\
INAF -- National Institute for Astrophysics, Trieste, Italy\\  
INFN -- National Institute for Nuclear Physics,
Sezione di Trieste, Italy
\and
D. Fabjan \at
Department of Astronomy, University of Trieste, via Tiepolo 11, I-34143 Trieste, Italy\\
\email{fabjan@oats.inaf.it}\\
INAF -- National Institute for Astrophysics, Trieste, Italy\\  
INFN -- National Institute for Nuclear Physics, Sezione di Trieste, Italy 
\and
L. Tornatore \at
Department of Astronomy, University of Trieste, via Tiepolo 11, I-34143 Trieste, Italy\\
\email{tornatore@oats.inaf.it}
\and
S. Schindler \at
Institut f\"ur Astro- und Teilchenphysik, Universit\"at
Innsbruck, Technikerstr. 25, 6020 Innsbruck, Austria\\
\email{Sabine.Schindler@uibk.ac.at}
\and
K. Dolag \at
                   Max-Planck-Institut f\"ur Astrophysik,
                  Karl-Schwarzschild Strasse 1, Garching bei
                  M\"unchen, Germany\\
\email{kdolag@mpa-garching.mpg.de}
\and
A. Diaferio \at
                  Dipartimento di Fisica Generale ``Amedeo Avogadro'',
                  Universit\`a degli Studi di Torino, Torino, Italy\\
                  \email{diaferio@ph.unito.it}\\ 
                   INFN -- National Institute for Nuclear Physics,
                  Sezione di Torino, Italy  
}

\date{Received: 1 October 2007 ; Accepted 7 November 2007}

\maketitle

\begin{abstract}
  The study of the metal enrichment of the intra--cluster and
  inter--galactic media (ICM and IGM) represents a direct means to
  reconstruct the past history of star formation, the role of feedback
  processes and the gas--dynamical processes which determine the
  evolution of the cosmic baryons. In this paper we review the
  approaches that have been followed so far to model the enrichment of
  the ICM in a cosmological context. While our presentation will be
  focused on the role played by hydrodynamical simulations, we will
  also discuss other approaches based on semi--analytical models of
  galaxy formation, also critically discussing pros and cons of the
  different methods. We will first review the concept of the model of
  chemical evolution to be implemented in any chemo--dynamical
  description. We will emphasise how the predictions of this model
  critically depend on the choice of the stellar initial mass
  function, on the stellar life--times and on the stellar yields. We
  will then overview the comparisons presented so far between X--ray
  observations of the ICM enrichment and model predictions. We will
  show how the most recent chemo--dynamical models are able to capture
  the basic features of the observed metal content of the ICM and its
  evolution. We will conclude by highlighting the open questions in
  this study and the direction of improvements for cosmological
  chemo--dynamical models of the next generation.
\end{abstract}

\section{Introduction}
\label{Introduction} 
Clusters of galaxies are the ideal cosmological signposts to trace the
past history of the inter--galactic medium (IGM), thanks to the high
density and temperature reached by the cosmic baryons trapped in their
gravitational potential wells (\citealt{2002ARA&A..40..539R}; \citealt{2005RvMP...77..207V};
\citealt{diaferio2008} - Chapter 2, this volume). Observations in the
X--ray band with the Chandra and XMM--Newton satellites are providing
invaluable information on the thermodynamical properties of the
intra--cluster medium (ICM) (\citealt{kaastra2008} - Chapter 9, this
volume). These observations highlight that non--gravitational sources
of energy, such as energy feedback from supernovae (SNe) and Active
Galactic Nuclei (AGN) have played an important role in determining the
ICM physical properties.  Spatially resolved X--ray spectroscopy
permits to measure the equivalent width of emission lines associated
to transitions of heavily ionised elements and, therefore, to trace
the pattern of chemical enrichment (e.g., 
\citealt{2004cgpc.symp..123M} for a review). In turn, this information
is inextricably linked to the history of formation and evolution of
the galaxy population (e.g., \citealt{1997ApJ...488...35R} and references therein), as inferred
from observations in the optical/near-IR band. For instance, \citet{2004A&A...419....7D} have first shown that cool core
clusters are characterised by a significant central enhancement of the
iron abundance, which closely correlates with the magnitude of the
Brightest Cluster Galaxies (BCGs). This demonstrates that a fully
coherent description of the evolution of cosmic baryons in the
condensed stellar phase and in the diffuse hot phase requires properly
accounting for the mechanisms of production and release of both energy
and metals.

The study of how these processes take place during the hierarchical
build up of cosmic structures has been tackled with different
approaches. Semi--analytical models (SAMs) of
galaxy formation provide a flexible tool to explore the space of
parameters which describe a number of dynamical and astrophysical
processes. In their most recent formulation, such models are coupled
to dark matter (DM) cosmological simulations, to trace the merging
history of the halos where galaxy formation takes place, and include a
treatment of metal production from type-Ia and type-II supernovae
(SN\,Ia and SN\,II, hereafter; \citealt{2004MNRAS.349.1101D}; 
\citealt{2005MNRAS.358.1247N}), so as to properly address the study of
the chemical enrichment of the ICM. The main limitation of this method
is that it does not include the gas dynamical processes which causes
metals, once produced, to be transported in the ICM. As a consequence,
they provide a description of the global metallicity of clusters and
its evolution, but not of the details of its spatial distribution.

In order to overcome this limitation, 
\citet{2006MNRAS.368.1540C} applied an alternative approach, in
which non--radiative hydrodynamical simulations of galaxy clusters are
used to trace at the same time the formation history of DM halos and
the dynamics of the gas. In this approach, metals are produced by SAM
galaxies and then suitably assigned to gas particles, thereby
providing a chemo--dynamical description of the ICM.  \citet{2006A&A...452..795D} used hydrodynamical simulations,
which include prescriptions for gas cooling, star formation and
feedback, similar to those applied in SAMs, to address the specific
role played by ram--pressure stripping to distribute metals, while
 \citet{2007A&A...466..813K} used the same approach
to study the different roles played by galactic winds and by
ram--pressure stripping. While these approaches offer advantages with
respect to standard SAMs, they still do not provide a fully
self--consistent picture, in which chemical enrichment is the outcome
of the process of star formation, associated to the cooling of the gas
infalling in high density regions. In this sense, a fully
self--consistent approach requires that the simulations include the
processes of gas cooling, star formation and evolution, along with the
corresponding feedback in energy and metals.

A number of authors have presented hydrodynamical simulations for the
formation of cosmic structures, which include treatments of the
chemical evolution at different levels of complexity.
\citet{1996A&A...315..105R} presented SPH simulations
of the Galaxy, forming in an isolated halo, by following iron and
oxygen production from SN\,II and SN\,Ia, also accounting for the effect
of stellar lifetimes. \citet{2001MNRAS.325...34M} 
performed a detailed analysis of chemo--dynamical SPH simulations,
aimed at studying both numerical stability of the results and the
enrichment properties of galactic objects in a cosmological context.
\citet{2002MNRAS.330..821L} discussed a statistical
approach to follow metal production in SPH simulations, which have a
large number of star particles, showing applications to simulations of
a disc--like galaxy and of a galaxy cluster. 
\citet{2003MNRAS.346..135K} carried out cosmological
chemo--dynamical simulations of elliptical galaxies, with an SPH code,
by including the contribution from SN\,Ia and SN\,II, also accounting for
stellar lifetimes. \citet{2003MNRAS.339.1117V} performed
an extended set of cluster simulations and showed that profiles of the
iron abundance are steeper than the observed ones. A similar analysis
has been presented by \citet{2006MNRAS.371..548R}, who
also considered the effect of varying the IMF and the feedback
efficiency on the enrichment pattern of the ICM. 
\citet{2005MNRAS.364..552S} presented an implementation of a model
of chemical enrichment in the {\tt GADGET-2} code, coupled to a
self--consistent model for star formation and feedback. In their
model, which was applied to study the enrichment of galaxies, they
included the contribution from SN\,Ia and SN\,II, assuming that all
long--lived stars die after a fixed delay time. 
\citet{Tornatore07,2004MNRAS.349L..19T} presented results
from an implementation of a detailed model of chemical evolution model
in the {\tt GADGET-2} code \citep{2005MNRAS.364.1105S}, including
metallicity--dependent yields and the contribution from intermediate
and low mass stars (ILMS hereafter).  The major advantage of this
approach is that the metal production is self--consistently predicted
from the rate of gas cooling treated by the hydrodynamical
simulations, without resorting to any external model. However, at
present the physical scales involved by the processes of star
formation and SN explosions are far from being resolved in simulations
which sample cosmological scales.  For this reason, such simulations
also need to resort to external sub--resolution models, which provide
an effective description of a number of relevant astrophysical
processes.

The aim of this paper is to provide a review of the results obtained
so far in the study of the chemical enrichment of the ICM in a
cosmological context. Although we will concentrate the discussion on
the results obtained from full hydrodynamical simulations, we will
also present results based on SAMs. As such, this paper complements
the reviews by \citealt{borgani2008} - Chapter 13, this volume, which
reviews the current status in the study of the thermodynamical
properties of the ICM with cosmological hydrodynamical simulations,
and by \citealt{schindler2008} - Chapter 17, this volume, which will
focus on the study of the role played by different physical processes
in determining the ICM enrichment pattern. Also, this paper  will not
present a detailed description of the techniques for simulations of
galaxy clusters, which is reviewed by \citealt{10_dolag2008} - Chapter 12,
 this volume.

In Sect.~\ref{chemical_evolution} we review the concept of model of
chemical evolution and highlight the main quantities which are
required to fully specify this model. In Sect.~\ref{globab} we
review the results on the global metal content of the ICM, while
Sect.~\ref{profiles} concentrates on the study of the metallicity
profiles and Sect.~\ref{evol} on the study of the ICM
evolution. Sect.~\ref{galaxies} discusses the properties of the galaxy population. Finally, Sect.~\ref{summary} provides a critical summary
of the results presented, by highlighting the open problems and lines
of developments to be followed by simulations of the next generation.

\section{What is a model of chemical evolution?}
\label{chemical_evolution}
In this section we provide a basic description of the key
ingredients required by a model of chemical evolution. For a more
detailed description we refer to the book by
\citet{2003ceg..book.....M}.

The process of star formation in cosmological hydrodynamical
simulations is described through the conversion of a gas element into
a star particle; as a consequence, in simulations of large volumes the
star particles have a mass far larger than that of a single star, with
values of the order of $10^6$--$10^8$~M$_\odot$, depending on the
resolution and on the mass of the simulated structure (e.g.,   \citealt{1996ApJS..105...19K}).  The consequence of this
coarse--grained representation of star formation is that each star
particle must be treated as a simple stellar population (SSP), i.e. as
an ensemble of coeval stars having the same initial metallicity. Every
star particle carries all the physical information (e.g. birth time
$t_{\mathrm b}$, initial metallicity and mass) that is needed to calculate
the evolution of the stellar populations that they represent, once
the lifetime function (see Sect.~\ref{lifet}), the IMF (see Sect.~\ref{imf}) and the yields (see Sect.~\ref{yields}) for SNe and ILMS
have been specified. Therefore, we can compute for every star particle
at any given time $t>t_{\mathrm b}$ how many stars are dying as SN\,II and SN\,Ia,
and how many stars undergo the AGB phase, according to the equations
of chemical evolution, that we will present here below.

The generation of star particles is pertinent to hydrodynamical
simulations, which include the description of the star formation. As
for semi--analytic models (SAMs) of galaxy formation, they generally
identify a galaxy with a suitably chosen DM particle in the
collisionless simulation which is used to reconstruct the merger
tree. While this DM particle defines the position of a galaxy, the
corresponding stellar content is described as a superposition of SSPs,
each generated in correspondence of the time step with which the
process of galaxy formation is studied (e.g.,  \citealt{2001MNRAS.323..999D};  \citealt{2004MNRAS.349.1101D}; 
 \citealt{2005MNRAS.358.1247N}, for applications of SAMs to the
cluster galaxies). In this sense, the description of the chemical
evolution model, as we provide here, can be used for both
hydrodynamical simulations and for SAMs.

It is generally assumed that the stars having mass above $8$~M$_\odot$ at
the end of the hydrostatic core burning undergo an electron capture
process, leading to a core collapse. A large amount of energy can be
transferred to the outer layers during this phase due to a substantial
production of neutrinos that easily escape from the central
core. Although theoretical work has not yet been able to reproduce a
sufficient energy deposition, it is currently supposed that this
process leads to an explosive ejection of the outer layers, giving
rise to a SN\,II. We remind the reader that $8$~M$_\odot$ is a commonly
adopted fiducial value, although the limiting mass for the onset of
explosive evolution is still debated (e.g.,  \citealt{1998A&A...334..505P}).

A different ejection channel is provided by the SN\,Ia that are believed
to arise from thermonuclear explosions of white dwarfs (WD hereafter)
in binary stellar systems as a consequence of the matter accretion
from the companion (e.g.,  \citealt{2000tias.conf...63N}). However, there are still a number of
uncertainties about the nature of both the WD and the companion and
about the mass reached at the onset of the explosion (e.g.,  \citealt{2001ApJ...558..351M}, 
\citealt{2000ApJ...528..108Y}). Furthermore, observational evidences for
multiple populations of Type Ia progenitors have been recently found
both from direct detection of Type Ia events (e.g.,  \citealt{2005A&A...433..807M}, 
\citealt{2005ApJ...629L..85S},  \citealt{2006ApJ...648..868S}) or from inferences on the chemical
enrichment patterns in galaxy clusters (e.g., 
\citealt{2004MNRAS.347..942G}).  Finally, a third way to eject heavy
elements in the interstellar medium is the mass loss of ILMS by
stellar winds.

In summary, the main ingredients that define a model of chemical
evolution are the following: {\em (a)} the SNe explosion rates, {\em
(b)} the adopted lifetime function, {\em (c)} the adopted yields and
{\em (d)} the IMF which fixes the number of stars of a given mass. We
describe each of these ingredients in the following.

\subsection{The equations of chemical evolution}

\subsubsection{Type Ia supernovae}
We provide  here below a short description of how Type Ia SN are
included in a model of chemical evolution. For a comprehensive review
of analytical formulations we refer to the paper by 
\citet{2005A&A...441.1055G}. 
Following  \citet{1983A&A...118..217G}, we assume
here that SN\,Ia arise from stars belonging to binary systems, having
mass in the range 0.8--8 M$_\odot$. Accordingly, in the
single--degenerate WD scenario 
 \citep{2000tias.conf...63N}, the rate of explosions of SN\,Ia can
be written as
\be
R_{{\mathrm{SN\,Ia}}}(t) \, = \,
A\,\int\limits_{\displaystyle{M_{\mathrm{B,inf}}}}^{\displaystyle{M_{\mathrm{B,sup}}}}
\phi(m_{\mathrm B})
\int\limits_{\displaystyle{\mu_{\mathrm m}}}^{\displaystyle{\mu_{\mathrm M}}} 
f(\mu)\,
\psi(t-\tau_{m_2})\,{\mathrm d}\mu\,{\mathrm d}m_{\mathrm B}\,. 
\label{eq:snia_rate}
\ee
In the above equation, $\phi(m)$ is the IMF, $m_{\mathrm B}$ is the total mass
of the binary system, $m_2$ is the mass of the secondary companion,
$\tau_m$ is the mass--dependent life--time
and $\psi(t)$ is the star formation
rate. The variable $\mu=m_2/m_{\mathrm B}$ is distributed according
to the function $f(\mu)$, while $A$ is the fraction of stars in binary
systems of that particular type to be progenitors of SN\,Ia (see
 \citealt{2001ApJ...558..351M} for more
details). For instance, in the model by 
\citet{1983A&A...118..217G} $\mu$ varies in the range between $\mu_{\mathrm m}$ and
$\mu_{\mathrm M}=0.5$, with $\mu_{\mathrm m}=\max\left[m_2(t)/m_{\mathrm B},(m_{\mathrm B}-0.5M_{\mathrm{BM}})/m_{\mathrm B}\right]$, where $m_2(t)$ is the
  mass of the companion which dies at the time $t$, according to the
  chosen life-time function.
Furthermore, let $M_{\mathrm{Bm}}$ and $M_{\mathrm{BM}}$ be the smallest and
largest value allowed for the progenitor binary mass $m_{\mathrm B}$. Then, the
integral over $m_{\mathrm B}$ runs in the range between $M_{\mathrm{B,inf}}$ and
$M_{\mathrm{B,sup}}$, which represent the minimum and the maximum value of the
total mass of the binary system that is allowed to explode at the time
$t$. These values in general are functions of $M_{\mathrm{Bm}}$, $M_{\mathrm{BM}}$, and
$m_2(t)$, which in turn depends on the star formation history
$\Psi(t)$. The exact functional dependence is defined by the SN\,Ia
progenitor model. For instance, in the model by 
\citet{1983A&A...118..217G} it is $M_{\mathrm{B,inf}} = \max[2m_2(t),
M_{\mathrm{Bm}}]$ and $M_{\mathrm{B,sup}} = 0.5 M_{\mathrm{BM}} + m_2(t)$. Under the assumption
of a short duration burst of star formation, the function $\psi(t)$
can be approximated with a Dirac $\delta$--function.  This case
applies to hydrodynamical simulations, which include star formation,
where the creation of a SSP is described by an impulsive star
formation event, while more complex descriptions should take into
account the continuous star formation history $\psi(t)$. Under the
above assumption for $\psi(t)$ and using the functional form of
$f(\mu)$ derived from statistical studies of the stellar population in
the solar neighbourhood 
\citep{1980IAUS...88...15T,2001ApJ...558..351M}, we find
\be
R_{\mathrm{SN\,Ia}}(t) \, = \, -{{\mathrm d}m_2(t)\over {\mathrm d}t}\bigg|_{m_2\equiv
  \tau^{-1}(t)}
24\,m_2^2\,A\int_{M_{\mathrm{Bm}}}^{M_{\mathrm{BM}}}\phi(m_{\mathrm B})\frac{1}{m_{\mathrm B}^3}{\mathrm d}m_{\mathrm B}\,.
\label{eq:GR83snia_rate}
\ee
Since the current understanding of the process of star formation does
not allow to precisely determine the value of $A$, its choice can be
fixed from the requirement of reproducing a specific observation, once
the form of the IMF is fixed. For instance, 
\citet{1995A&A...304...11M} found that $A=0.1$ was required to
reproduce the observed iron enrichment.

\subsubsection{Supernova Type II and low and intermediate mass
  stars} 
Computing the rates of SN\,II and ILMS is conceptually
simpler, since they are driven by the lifetime function $\tau(m)$
convolved with the star formation history $\psi(t)$ and multiplied
by the IMF $\phi(m=\tau^{-1}(t))$. Again, since 
$\psi(t)$ is a delta--function for the SSP used in
simulations, 
the SN\,II and ILMS rates read 
\be
R_{{\mathrm SN\,II}|{\mathrm ILMS}}(t)=\phi(m(t)) \times \left( -\frac{{\mathrm d}\,m(t)}{{\mathrm d}\, t}\right)
\label{eq:my_snii_rate}
\ee
where $m(t)$ is the mass of the star that dies at time $t$.  We note
that the above expression must be multiplied by a factor of $(1-A)$
for AGB rates if the interested mass $m(t)$ falls in the same range of
masses which is relevant for the secondary stars of SN\,Ia binary
systems.

In order to compute the metal release
by stars (binary systems in case of SN\,Ia) of a given mass we
need to take into account the yields $p_{Z_i}(m, Z)$, which provide the
mass of the element $i$ produced by a star of mass $m$ and initial
metallicity $Z$. Then, the equation which describes the evolution of
the mass $\rho_i(t)$ for the element $i$, holding for a generic form
of the star formation history $\psi(t)$, reads: 
\be
\begin{array}{lcl}
\dot{\rho}_i(t) &= &-\psi(t)Z_i(t) \\
&\\
& + & A\int\limits_{M_{\mathrm{Bm}}}^{M_{\mathrm{BM}}}\phi(m)
\left[\int\limits_{\mu_{\min}}^{0.5}
f(\mu)\psi(t-\tau_{m_2})p_{Z_i}(m, Z)\,{\mathrm d}\mu \right]\, {\mathrm d}m  \\
&\\
& + & (1-A)\int\limits_{M_{\mathrm{Bm}}}^{M_{\mathrm{BM}}}
\psi(t-\tau(m))p_{Z_i}(m, Z)\varphi(m)\,{\mathrm d}m  \\
&\\
& + & \int\limits_{M_{\mathrm L}}^{M_{\mathrm{Bm}}}
\psi(t-\tau(m))p_{Z_i}(m, Z)\varphi(m)\,{\mathrm d}m  \\
&\\
& + & \int\limits_{M_{\mathrm{BM}}}^{M_{\mathrm{U}}}
\psi(t-\tau(m))p_{Z_i}(m, Z)\varphi(m)\,{\mathrm d}m.
\label{eq:stev}
\end{array}
\ee 
In the above equation, $M_{\mathrm L}$ and $M_{\mathrm U}$ are the minimum and maximum
mass of a star, respectively. Commonly adopted choices for these
limiting masses are $M_{\mathrm L}\simeq 0.1$~M$_\odot$ and $M_{\mathrm U}\simeq
100$~M$_\odot$. The term in the first line of Eq.~\ref{eq:stev} accounts
for the metallicity sink due to the locking of metals in the newborn
stars. The term in the second line accounts for metal ejection
contributed by SN\,Ia.  The terms in the third and fourth lines describe
the enrichment by mass--loss from intermediate and low mass stars,
while the last line accounts for ejecta by SN\,II.

\subsection{The lifetime function}
\label{lifet}
Different choices for the mass--dependence of the life--time function
have been proposed in the literature. For instance, \citet{1993ApJ...416...26Pb} (PM93 hereafter) proposed the expression
\be
\label{eq:PM_lifetimes}
\tau(m) = \left\{\begin{array}{ll}
10^{[(1.34 - \sqrt{1.79 - 0.22 (7.76 -\log(m))}) / 0.11 ]
  - 9 } \,\,\,
{\rm for}\,\,m \le 6.6~{\rm M}_\odot\\
\\
1.2\, m^{-1.85} + 0.003 \,\,\,\,\, {\rm otherwise.} \\
\end{array}\right.
\ee
An alternative expression has been proposed by 
\citet{1989A&A...210..155M} (MM89 hereafter), and extrapolated by
 \citet{1997ApJ...477..765C} to very high ($> 60$~M$_\odot$) and very low ($< 1.3$~M$_\odot$) masses:
\be
\label{eq:MM_lifetimes}
\tau(m)=\left\{\begin{array}{ll}
10^{ -0.6545 \log m + 1} & m \le 1.3~{\rm M}_\odot\\
\vspace{-0.2cm}&\\
10^{ -3.7 \log m + 1.351} & 1.3 < m \le 3~{\rm M}_\odot\\
\vspace{-0.2cm}&\\
10^{ -2.51 \log m + 0.77} & 3 < m \le 7~{\rm M}_\odot\\
\vspace{-0.2cm}&\\
10^{ -1.78 \log m + 0.17} & 7 < m \le 15~{\rm M}_\odot\\
\vspace{-0.2cm}&\\
10^{ -0.86 \log m - 0.94} & 15 < m \le 53~{\rm M}_\odot\\
\vspace{-0.3cm}&\\
1.2 \times m^{-1.85}+ 0.003 & {\rm otherwise.} 
\end{array} \right.
\ee

The main difference between these two functions concerns the
life--time of low mass stars ($ < 8\,$M$_\odot$). The MM89 life--time
function delays the explosion of stars with mass $\gtrsim
1~$M$_\odot$, while it anticipates the explosion of stars below
$1~$M$_\odot$ with respect to the PM93 life--time function. Only for
masses below $1$~M$_\odot$ does the PM93 function predict much more
long--living stars.  This implies that different life--times will
produce different evolution of both absolute and relative abundances
(we refer to  \citet{2005A&A...430..491R} for a detailed
description of the effect of the lifetime function in models of
chemical evolution).

\begin{figure}
\centerline{
\hbox{
\includegraphics[width=0.49\textwidth]{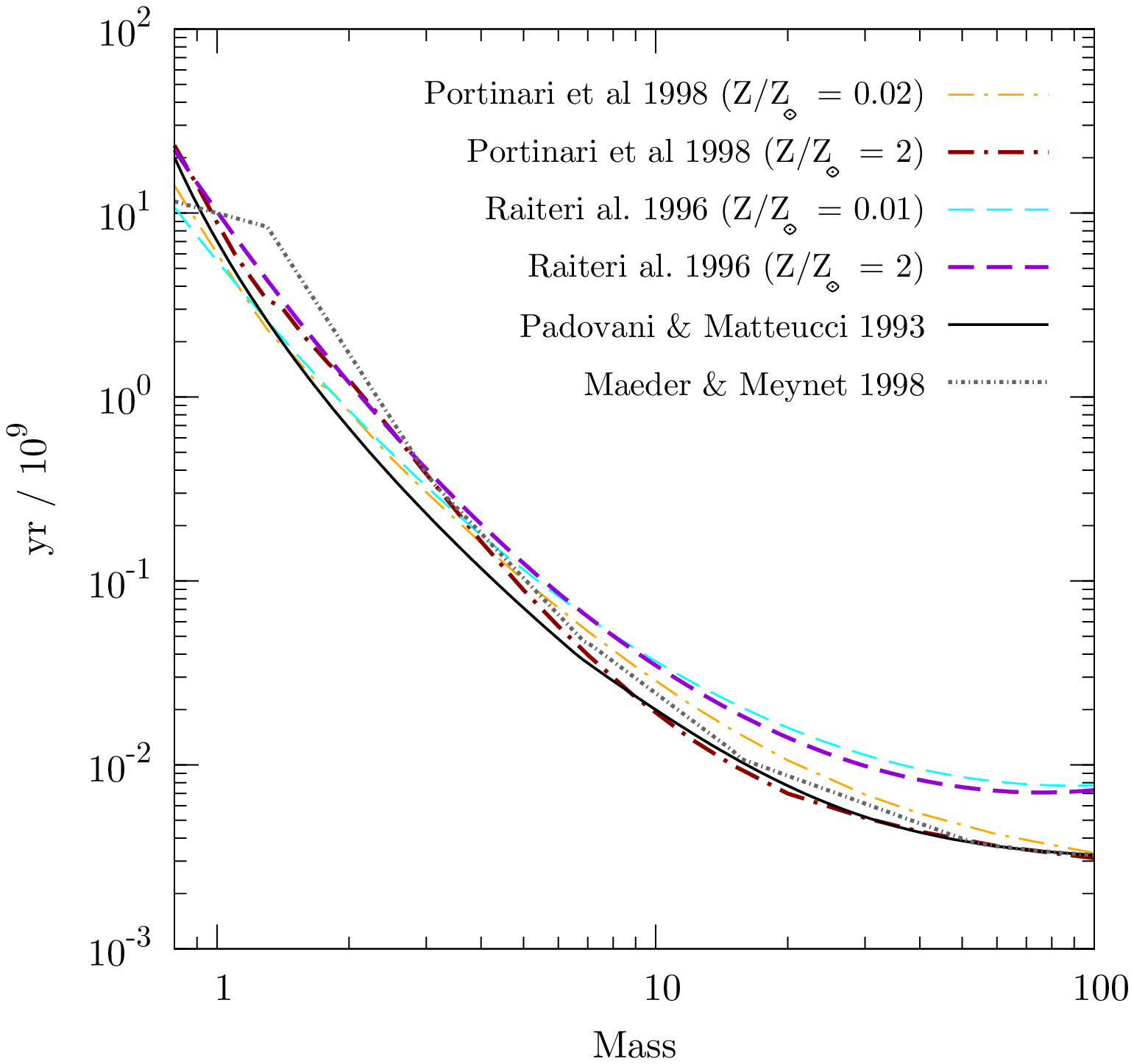}
\includegraphics[width=0.49\textwidth]{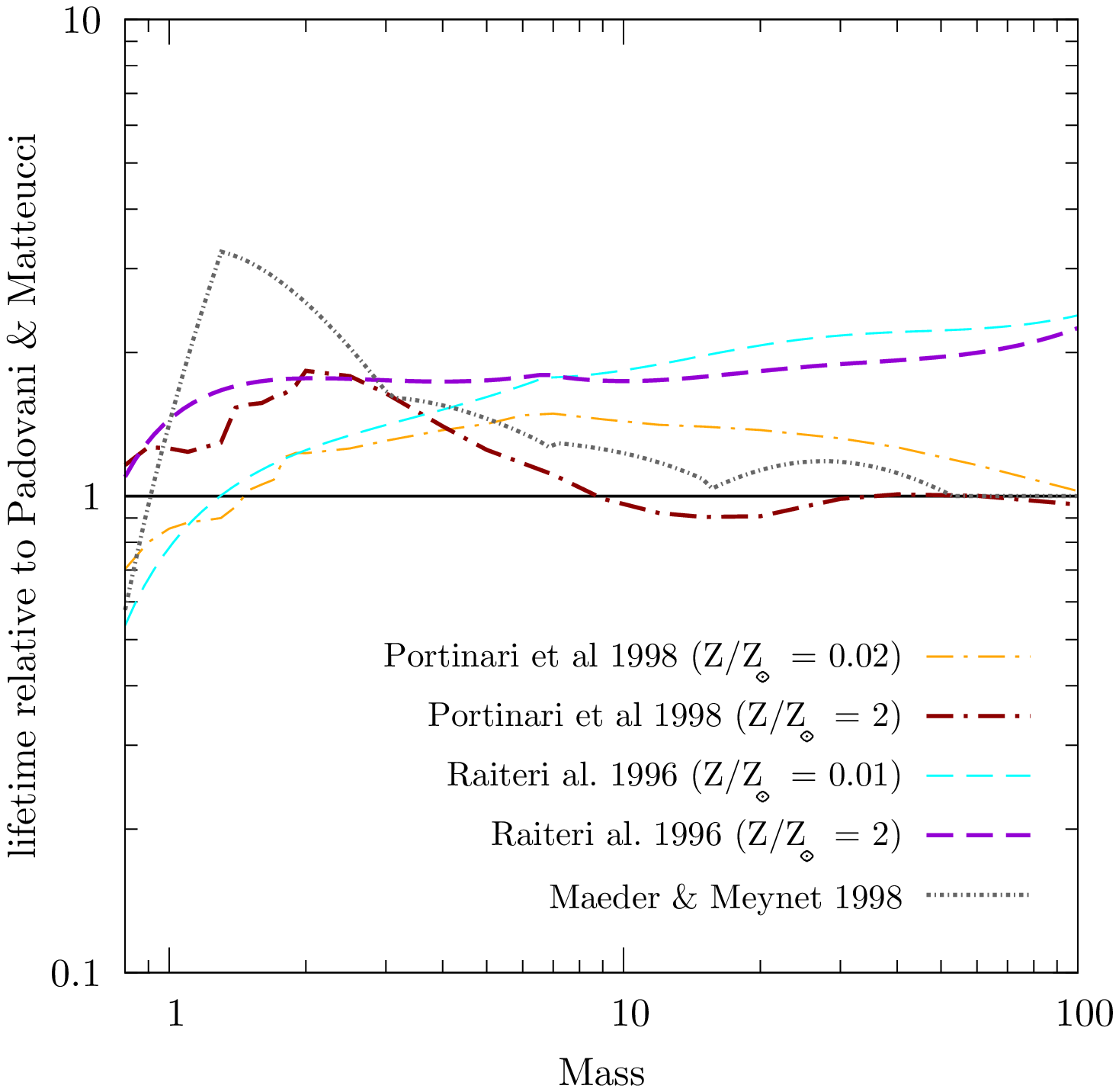} 
}}
\caption{Left panel: the mass dependence of different lifetime
  functions: Padovani \&  Matteucci (1993, solid line), Maeder
  \&  Meynet (1998, dotted line), Raiteri et al. (1996, dashed lines;
  the thin line refers to $Z=0.02\,Z_\odot$, the thick line refers to
  $Z=2\,Z_\odot$), Portinari et al. (1998, dot--dashed lines; the thin
  line refers to $Z=0.02\,Z_\odot$, the thick line refers to
  $Z=2\,Z_\odot$). Right panel: the ratio between the same lifetimes
  and that by Padovani \&  Matteucci (1993).}
\label{fi:lifetimes}
\end{figure}

We point out that the above lifetime functions are independent of
metallicity, whereas in principle this dependence can be included in a
model of chemical evolution (e.g.,  \citealt{1998A&A...334..505P}). For instance,  \citet{1996A&A...315..105R} used the metallicity--dependent
lifetimes as obtained from the Padova evolutionary tracks
\citep{1994A&AS..106..275B}. We show in Fig.~\ref{fi:lifetimes} a
comparison of different lifetime functions.

\subsection{Stellar yields}
\label{yields}

The stellar yields specify the quantity $p_{Z_i}(m, Z)$, which appears
in Eq.~\ref{eq:stev} and, therefore, the amount of different metal
species which are released during the evolution of a SSP. A number of
different sets of yields have been proposed in the literature, such as
those by  \citet{1981A&A....94..175R},  \citet{1997A&AS..123..305V}, 
\citet{2001A&A...370..194M} for the LIMS and those by  \citet{1997NuPhA.621..467N},  \citet{1999ApJS..125..439I},  \citet{2003NuPhA.718..139T} for SN\,Ia.  As for SN\,II, there are
many proposed sets of metallicity--dependent yields; among others,
those by \citet{1995ApJS..101..181W}, by
 \citet{1998A&A...334..505P}, by  \citet{2004ApJ...608..405C}, which are based on different
assumptions of the underlying model of stellar structure and
evolution.

As an example of the differences among different sets of yields, we
show in the left panel of Fig.~\ref{fi:imfs} the ratios between the
abundances of different elements produced by the SN\,II of a SSP, as
expected from  \citet{1995ApJS..101..181W} and from
 \citet{2004ApJ...608..405C}. Different curves and
symbols here correspond to different values of the initial metallicity
of the SSP. Quite apparently, the two sets of yields provide
significantly different metal masses, by an amount which can
sensitively change with initial metallicity. This illustrates how a
substantial uncertainty exists nowadays about the amount of metals
produced by different stellar populations. There is no doubt that
these differences between different sets of yields represent one of
the main uncertainties in any modelling of the chemical evolution
of the ICM.

For this reason, we recommend that papers related to the cosmological
modelling of the chemical evolution, either by SAMs or by
hydrodynamical simulations, should be very careful in specifying for
which set of yields the computations have been performed as well as
the details of assumed lifetime and models for SN\,II and SN\,Ia.
In the absence of this, it becomes quite hard to judge the reliability
of any detailed comparison with observational data or with the
predictions of other models.

\subsection{The initial mass function}
\label{imf}
The initial mass function (IMF) is one of the most important quantities
in a model of chemical evolution. It directly determines the
relative ratio between SN\,II and SN\,Ia and, therefore, the relative
abundance of $\alpha$--elements and Fe--peak elements.  The shape of
the IMF also determines how many long--living stars will form with
respect to massive short--living stars. In turn, this ratio affects
the amount of energy released by SNe and the present luminosity of
galaxies, which is dominated by low mass stars, and the (metal)
mass--locking in the stellar phase.

As of today, no general consensus has been reached on whether the IMF
at a given time is universal or strongly dependent on the environment,
or whether it is time--dependent, i.e. whether local variations of the
values of temperature, pressure and metallicity in star--forming
regions affect the mass distribution of stars.

The IMF $\phi(m)$ is defined as the number of stars of a given mass per
unit logarithmic mass interval. A widely used form is
\be
\phi(m)\,=\,dN/d\log m \propto m^{-x(m)}\,. 
\ee 
If the exponent $x$ in the above expression does not depend on the
mass $m$, the IMF is then described by a single power--law. The most
famous and widely used single power--law IMF is the 
\citet{1955ApJ...121..161S} one that has $x=1.35$.  
\citet[AY hereafter]{1987A&A...173...23A} proposed an
IMF with $x=0.95$, which predicts a relatively larger number of
massive stars. In general, IMFs providing a large number of massive
stars are usually called top--heavy. 
More recently, different expressions of the IMF have been proposed in
order to model a flattening in the low--mass regime that is currently
favoured by a number of observations.
\citet{2001MNRAS.322..231K} introduced a
multi--slope IMF, which is defined as
\be
\phi(m)\, \propto \,\left\{\begin{array}{ll}
m^{-1.3} & \qquad m \ge 0.5\,{\rm M}_\odot\\ 
\vspace{-0.2cm}&\\
m^{-0.3} & \qquad 0.08 \le m < 0.5\,{\rm M}_\odot\\ 
\vspace{-0.2cm}&\\
m^{\ 0.7} & \qquad m \le 0.08\,{\rm M}_\odot\\ 
\end{array} \right.
\label{Kroupa}
\ee 
\citet{2003PASP..115..763C} proposed another expression for
the IMF, which is quite similar to that one proposed by Kroupa
\be
\phi(m)\, \propto \, \left\{\begin{array}{ll}
m^{-1.3} & \qquad m > 1\, {\rm M}_\odot\\
\vspace{-0.2cm}&\\
e^{\frac{-\left(\log(m) -
        \log(m_c)\right)^2}{2\, \sigma^2}} & \qquad m \le 1\, {\rm
  M}_\odot\\
\end{array}\right.
\label{Chabrier}
\ee
Theoretical arguments (e.g.  \citealt{1998MNRAS.301..569L})
suggest that the present--day characteristic mass scale, where the IMF
changes its slope, $\sim 1$~M$_\odot$ should have been larger in the
past, so that the IMF at higher redshift was top--heavier than at
present.

While the shape of the IMF is determined by the local conditions of
the inter--stellar medium, direct hydrodynamical simulations of star
formation in molecular clouds are only now approaching the required
resolution and sophistication level to make credible predictions on
the IMF (e.g.,  \citealt{2006MNRAS.368.1296B},
 \citealt{2007ApJ...661..972P}; see 
\citealt{2007arXiv0707.3514M} for a detailed discussion).

\begin{figure}
\centerline{
\hbox{
\includegraphics[width=0.49\textwidth]{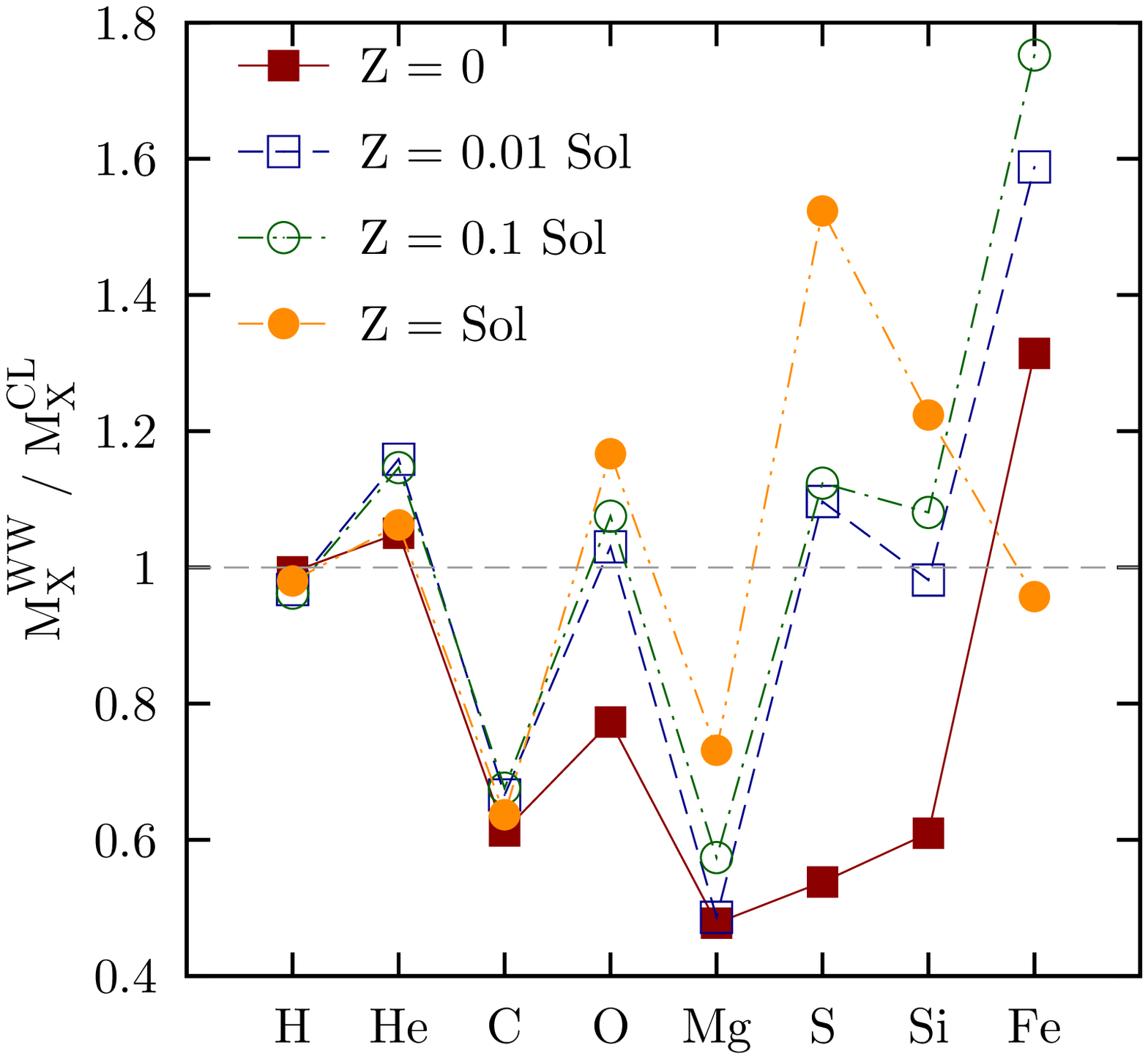}
\includegraphics[width=0.49\textwidth]{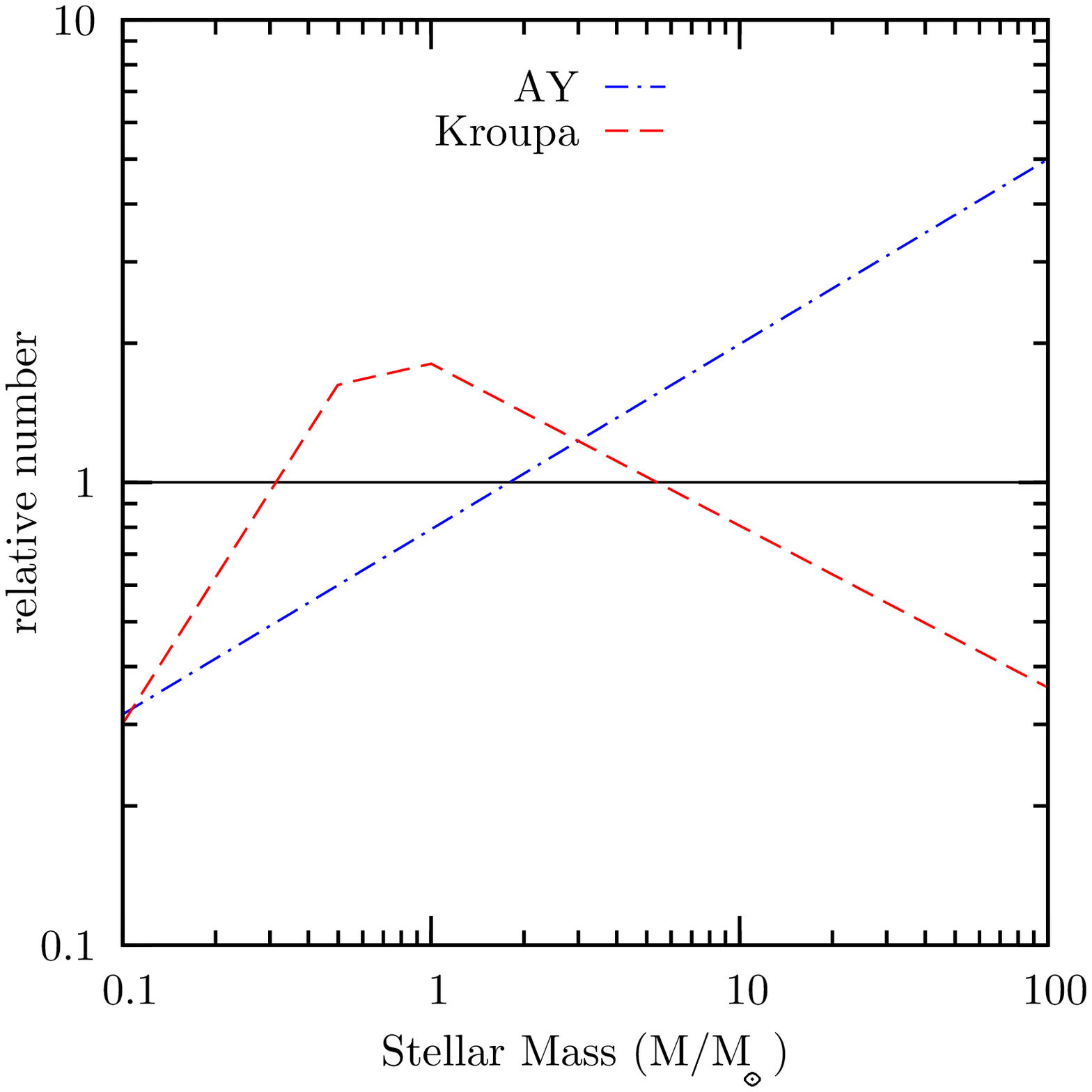} 
}}
\caption{Left panel: the ratio $M_j^{\mathrm{WW}}/M_j^{\mathrm{CL}}$ between the
  mass of species $j$, produced by the SN\,II of a SSP, when using the
  two sets of yields by 
  \protect\citet{1995ApJS..101..181W} and by 
  \protect\citet{2004ApJ...608..405C} for different values of the
  initial SSP metallicity. Different symbols are for different values
  of the initial metallicity of the SSP, as reported by the labels.
  From  \protect\citet{Tornatore07}. Right
  panel: the shape of different IMFs introduced in the
  literature. Stellar masses are expressed in units of M$_\odot$.}
\label{fi:imfs}
\end{figure}

We show in Fig.~\ref{fi:imfs} the number of stars, as a function of
their mass, predicted by different IMFs, relative to those of the
Salpeter IMF. As expected, the AY IMF predicts a larger number of
high--mass stars and, correspondingly, a smaller number of low--mass
stars, the crossover taking place at $\simeq 2$~M$_\odot$. As a result,
we expect that the enrichment pattern of the AY IMF will be
characterised by a higher abundance of those elements, like oxygen,
which are mostly produced by SN\,II.  Both the Kroupa and the Chabrier
IMFs are characterised by a relative deficit of very low--mass stars
and a mild overabundance of massive stars and ILMS. Correspondingly, an
enhanced enrichment in both Fe--peak and $\alpha$ elements like oxygen is
expected, mostly due to the lower fraction of mass locked in
ever--living stars. For reference we also show a different IMF, also
proposed by  \citet{1993MNRAS.262..545K}, that exhibits
a deficit in both very low-- and high--mass stars; for this kind of
IMF a lower $\alpha\,/\,$Fe ratio with respect to the Salpeter IMF is
expected.

Since clusters of galaxies basically behave like ``closed boxes'', the
overall level of enrichment and the relative abundances should
directly reflect the condition of star formation. While a number of
studies have been performed so far to infer the IMF shape from the
enrichment pattern of the ICM, no general consensus has been
reached. For instance, 
\citet{1997ApJ...488...35R,2004cgpc.symp..260R} and  \citet{2002NewA....7..227P} argued that both the global level
of ICM enrichment and the $\alpha$/Fe relative abundance can be
reproduced by assuming a Salpeter IMF, as long as this relative
abundance is sub-solar. Indeed, a different conclusion has been
reached by other authors in the attempt of explaining the relative abundances
$[\alpha/{\rm Fe}]\,\simg 0$ in the ICM (e.g.,  \citealt{2005ApJ...620..680B}).  \citet{2004ApJ...604..579P} used a phenomenological model to
argue that a standard Salpeter IMF cannot account for the observed
$\alpha$/Fe ratio in the ICM. As we shall discuss in the following, a
similar conclusion was also reached by  \citet{2005MNRAS.358.1247N}, who used semi--analytical models of
galaxy formation to trace the production of heavy elements, and by
 \citet{2005MNRAS.361..983R}, who used hydrodynamical
simulations including chemical enrichment.  \citet{2006MNRAS.373..397S} analysed the galaxy population from
simulations of galaxy clusters and concluded that a Salpeter IMF
produces a colour--magnitude relation that, with the exception of the
BCGs, is in reasonable agreement with observations. On the contrary,
the stronger enrichment provided by a top--heavier IMF turns into too
red galaxy colours.

In the following sections, the above results on the cosmological
modelling of the ICM will be reviewed in more detailed, also
critically discussing the possible presence of observational biases,
which may alter the determination of relative abundances and,
therefore, the inference of the IMF shape.

Different variants of the chemical evolution model have been
implemented by different authors in their simulation codes.  For
instance, the above described model of chemical evolution has been
implemented with minimal variants by 
\citet{2003MNRAS.340..908K}, 
\citet{2004MNRAS.347..740K}, who also included the effect of
hypernova explosions, and  \citet{Tornatore07}, while
 \citet{2007arXiv0707.1433T} also included the
effect of metal enrichment from low--metallicity (Pop~III)
stars.  \citet{1996A&A...315..105R} and 
\citet{2003MNRAS.339.1117V} also used a similar model, but neglected
the contribution from low- and intermediate-mass stars.  \citet{2001MNRAS.325...34M} and  \citet{2005MNRAS.364..552S} neglected delay times for SN\,II,
assumed a fixed delay time for SN\,Ia and neglected the contribution to
enrichment from low- and intermediate-mass stars.

Clearly, a delicate point in hydrodynamical simulations is deciding
how metals are distributed to the gas surrounding the star particles.
The physical mechanisms actually responsible for enriching the
inter--stellar medium (ISM; e.g., stellar winds, blast waves from SN
explosions, etc.) take place on scales which are generally well below
the resolution of current cosmological simulations of galaxy clusters
(see \citealt{schindler2008} - Chapter 17, this volume).
For this reason, the usually adopted procedure is that of distributing
metals according to the same kernel which is used for the computation
of the hydrodynamical forces, a choice which is anyway quite
arbitrary.  \citet{2001MNRAS.325...34M} and  \citet{Tornatore07} have tested the effect of changing in
different ways the weighting scheme to distribute metals and found
that final results on the metal distribution are generally rather
stable.  Although this result is somewhat reassuring, it is clear that
this warning on the details of metal distribution should always be
kept in mind, at least until our simulations will have enough
resolution and accurate description of the physical processes
determining the ISM enrichment (\citealt{schindler2008} - Chapter 17,
this volume).

\begin{figure*}
\centerline{
\hbox{
\includegraphics[width=0.49\textwidth]{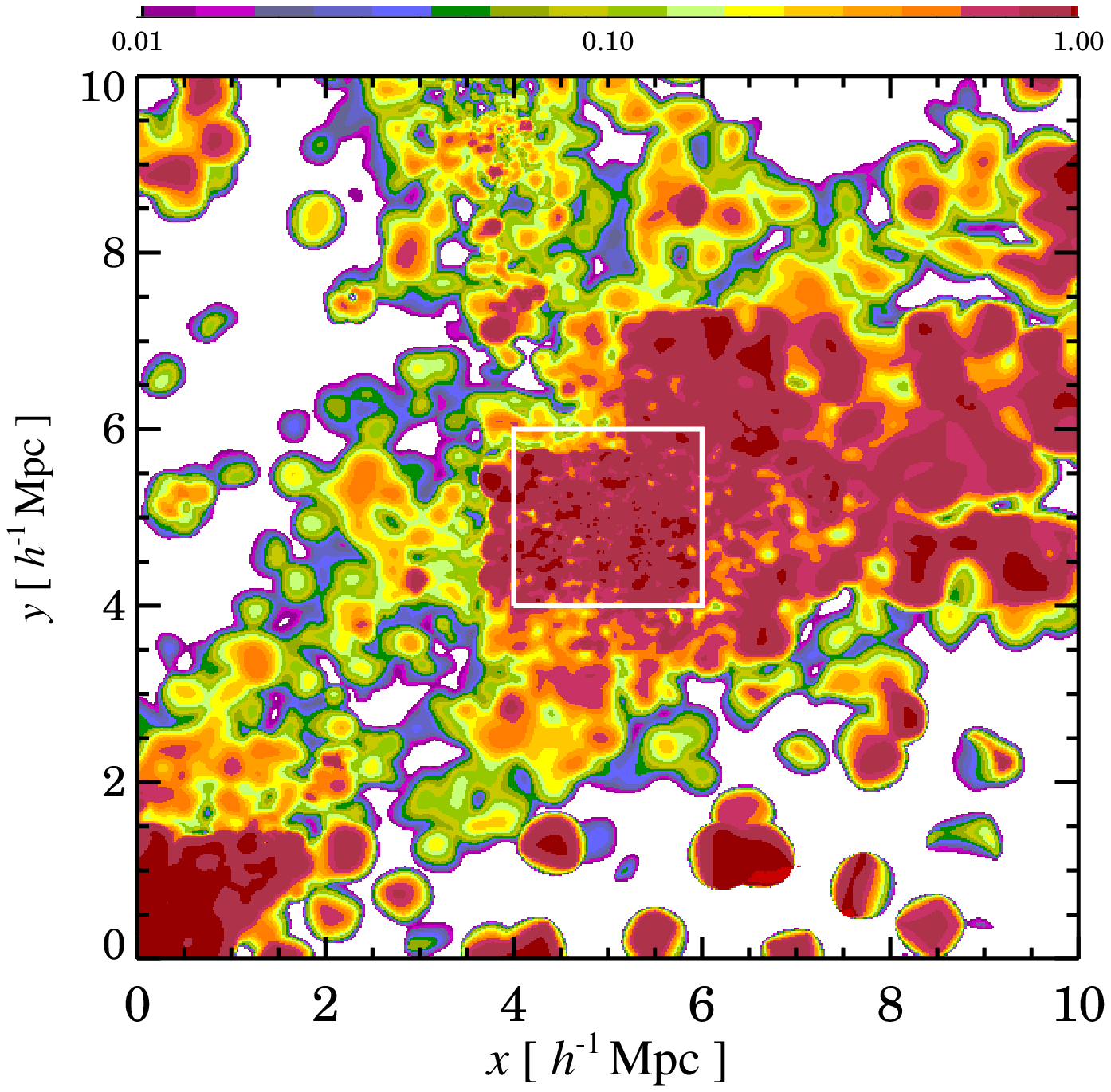} 
\includegraphics[width=0.49\textwidth]{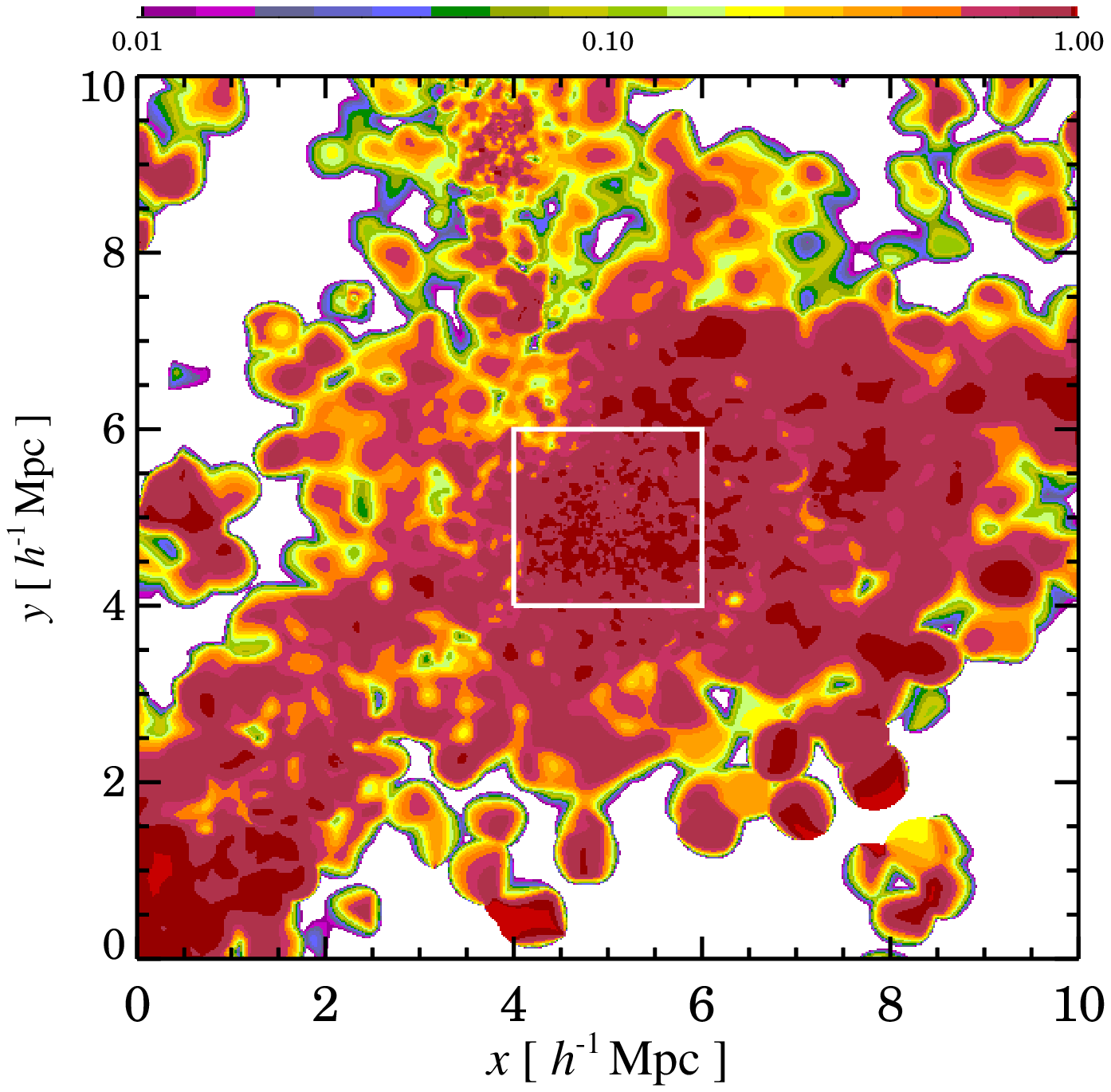}
}}
\caption{Maps of the fractional contribution of SN\,II to the global
  metallicity (from  \protect\citealt{Tornatore07}). The left
  panel refer to the Salpeter IMF, while the right panel is for the AY
  top--heavy IMF. The white square is centred on the minimum of the
  cluster gravitational potential and its size is about twice the
  value of $R_{200}$. }
\label{fi:maps_imf} 
\end{figure*}

\begin{figure*}
\centerline{
\includegraphics[width=0.7\textwidth]{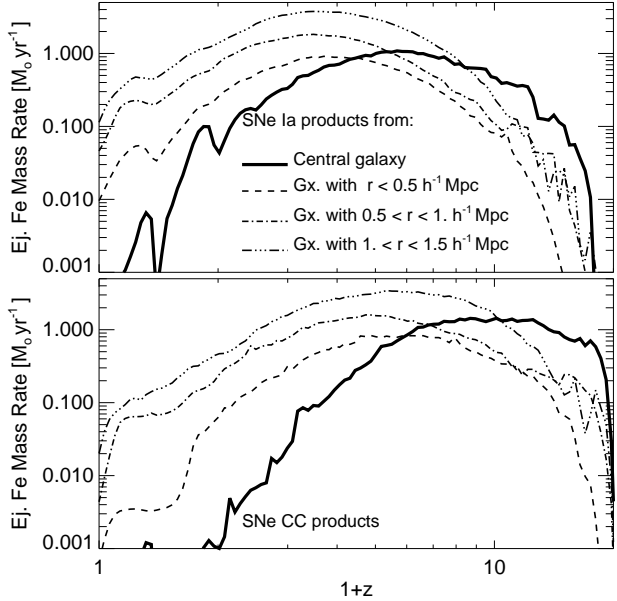} 
}
\caption{Iron mass ejection rate as a function of redshift for SN\,Ia
  (top panel) and SN\,II (bottom panel), as predicted by the
  semi--analytical model by 
  \protect\citet{2006MNRAS.368.1540C}. The thick solid line is for
  the products released by the BCG, while the other curves are for
  galaxies lying within different cluster-centric distances, as
  computed at $z=0$, after
  excluding the BCG contribution.}
\label{fi:enrich_imf} 
\end{figure*}

\section{Global abundances}
\label{globab}
As a first step in our overview of the results of the ICM enrichment,
we discuss a few general features of the enrichment pattern, which are
related to the underlying model of chemical evolution. As a first
example, we show in Fig.~\ref{fi:maps_imf} maps of the fractional
contribution of SN\,II to the global metallicity obtained from SPH
simulations of a galaxy cluster having a temperature $\simeq 3$ keV
\citep{Tornatore07}; the simulations include a self--consistent
description of gas cooling, star formation and chemical evolution. The
left and the right panels show the results based on assuming a
Salpeter and a top--heavy AY IMFs, respectively. The clumpy aspect of
these maps reveal that the products of SN\,II and SN\,Ia are spatially
segregated. Since SN\,II products are released over a relatively short
time--scale, they are preferentially located around star forming
regions. On the other hand, SN\,Ia products are released over a longer
time scale, which is determined by the life--times of the
corresponding progenitors. As a consequence, a star particle has time
to move away from the region, where it was formed, before SN\,Ia
contribute to the enrichment. In this sense, the different spatial
pattern of SN\,Ia and SN\,II products also reflects the contribution that
a diffuse population of intra--cluster stars (e.g., 
\citet{2004IAUS..217...54A} for a review) can provide to the ICM
enrichment. Also, using a top--heavier IMF turns into a larger number
of SN\,II (see also Fig. \ref{fi:imfs}), with a resulting increase of
their relative contribution to the metal production.

To further illustrate the different timing of the enrichment from SN\,Ia
and SN\,II, we show in Fig.~\ref{fi:enrich_imf} the rate of the iron
mass ejection from the two SN populations, as predicted by a SAM model
of galaxy formation, coupled to a non--radiative hydrodynamical
simulation \citep{2006MNRAS.368.1540C}. As expected, the contribution
from SN\,II (lower panel) peaks at higher redshift than that of the
SN\,Ia, with a quicker decline at low $z$, and closely traces the star
formation rate (SFR). As for the SN\,Ia, their rate of enrichment is
generally given by the convolution of the SFR with the lifetime
function. As such, it is generally too crude an approximation to
compute the enrichment rate by assuming a fixed delay time with
respect to the SFR. Also interesting to note from this figure is that
everything takes places earlier in the BCG (solid curves), than in
galaxies at larger cluster-centric distances.

A crucial issue when performing any such comparison between
observations and simulations concerns the definition of
metallicity. From observational data, the metallicity is computed
through a spectral fitting procedure, by measuring the equivalent
width of an emission line associated to a transition between two
heavily ionised states of a given element. In this way, one expects
that the central cluster regions, which are characterised by a
stronger emissivity, provides a dominant contribution to the global
spectrum and, therefore, to the observed ICM metallicity. The simplest
proxy to this spectroscopic measure of the ICM metallicity is,
therefore, to use the emission--weighted definition of metallicity:
\be 
Z_{\mathrm{ew}}\,=\,{\sum_i Z_i m_i \rho_{g,i}\Lambda(T_i,Z_i)\over
  \sum_i m_i \rho_{g,i}\Lambda(T_i)}\,.
\label{eq:z_ew}  
\ee
In the above equation, $Z_i$, $m_i$, $\rho_{g,i}$ and $T_i$ are the
metallicity, mass, density and temperature of the $i$--gas element,
with the sum being performed over all the gas elements lying within
the cluster extraction region. Furthermore, $\Lambda(T,Z)$ is the
metallicity dependent cooling function, which is defined so that $n_{\mathrm e}
n_{\mathrm H}\Lambda(T,Z)$ is the radiated energy per unit time and unit volume
from a gas element having electron number density $n_{\mathrm e}$ and hydrogen
number density $n_{\mathrm H}$ (e.g., 
\citealt{1993ApJS...88..253S}). Since both simulated and observed
metallicity radial profiles are characterised by significant negative
gradients, we expect the ``true'' mass--weighted metallicity,
\be 
Z_{\mathrm{mw}}\,=\,{\sum_i Z_i m_i\over \sum_i m_i}\,,
\label{eq:z_mw}  
\ee
to be lower than the ``observed'' emission--weighted estimate. Indeed,
mock spectral observations of simulated clusters have shown that the
above emission--weighted definition of metallicity is generally quite
close to the spectroscopic value  \citep{2007arXiv0707.1573K,2007arXiv0707.2614R}, at least for iron. On the other hand,
Rasia et al. have also shown that abundances of other elements may be
significantly biased. This is especially true for those elements, like
oxygen, whose abundance is measured from transitions taking place at
energies at which it is difficult to precisely estimate the level of the
continuum in a hot system, due to the limited spectral resolution
offered by the CCD on-board Chandra and XMM--Newton.

\begin{figure}
\centerline{
\includegraphics[width=\textwidth]{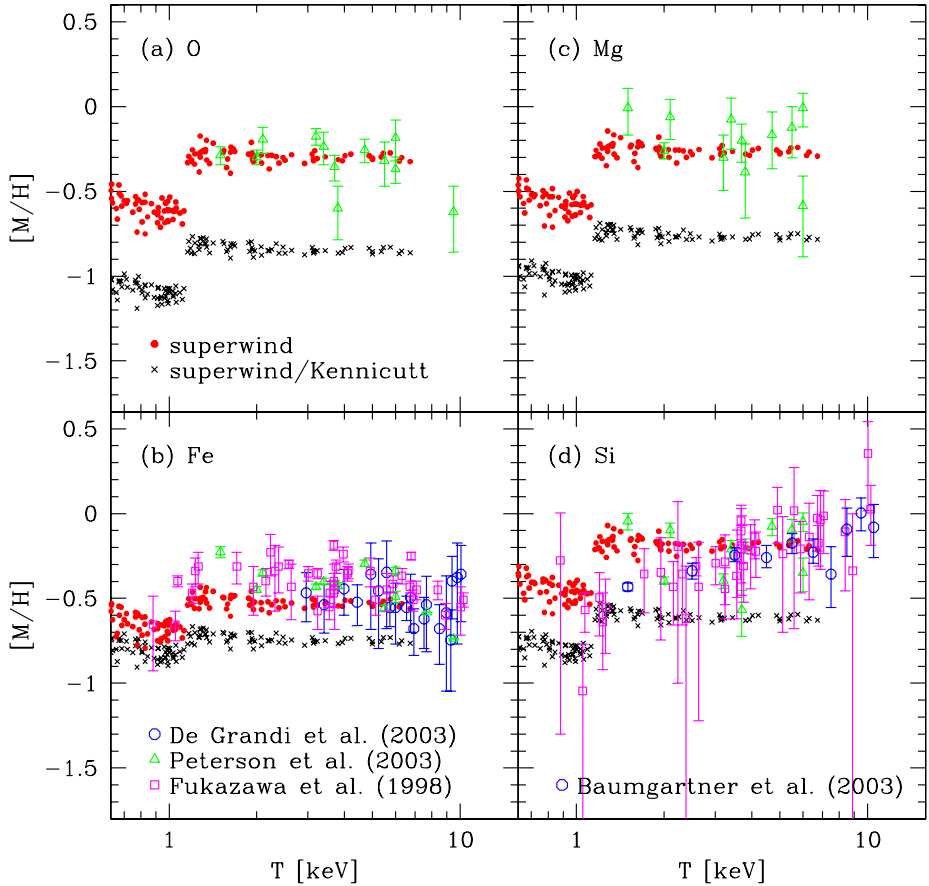}}
\caption{The metal abundances of the ICM, as predicted by a
  semi--analytical model of galaxy formation (from  \protect\citealt{2005MNRAS.358.1247N}). Each panel shows a
  different element: (a) [O/H], (b) [Fe/H], (c) [Mg/H] and (d) [Si/H].
  Model predictions are shown by dots for a top-heavy IMF, and by
  crosses for the Kennicutt IMF. Symbols with error bars correspond to
  observational data, which have been taken from  \protect\citet{2004A&A...419....7D},  \protect\citet{2003ApJ...590..207P},  \protect\citet{1998PASJ...50..187F} and  \protect\citet{2005ApJ...620..680B} (see  \protect\citet{2005MNRAS.358.1247N} for details). All
  abundances have been rescaled to the Solar abundances by  \protect\citet{1998SSRv...85..161G}.}
\label{fi:fig_naga}
\end{figure}

\begin{figure}
\hbox{
\includegraphics[width=0.49\textwidth]{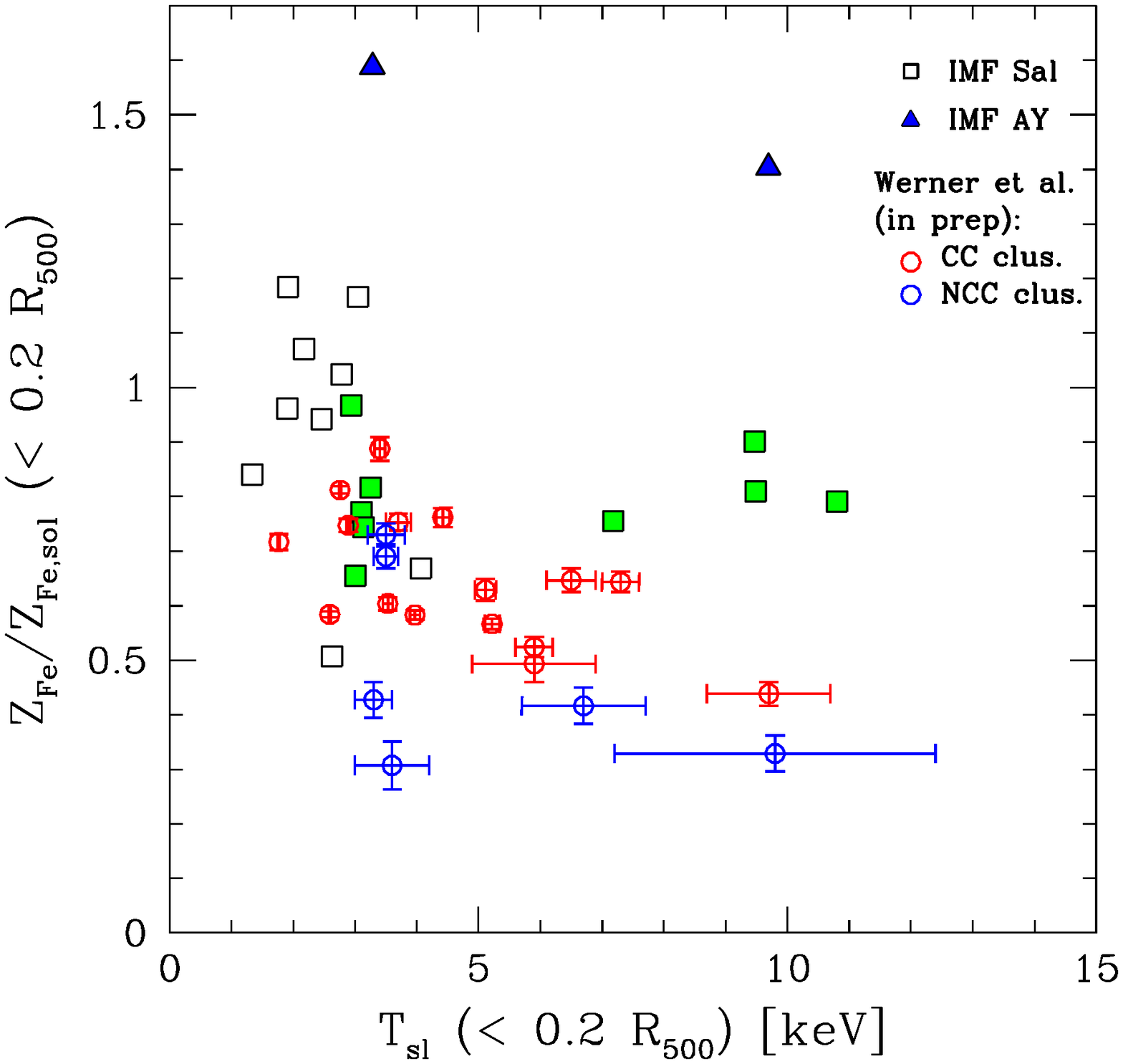}
\includegraphics[width=0.49\textwidth]{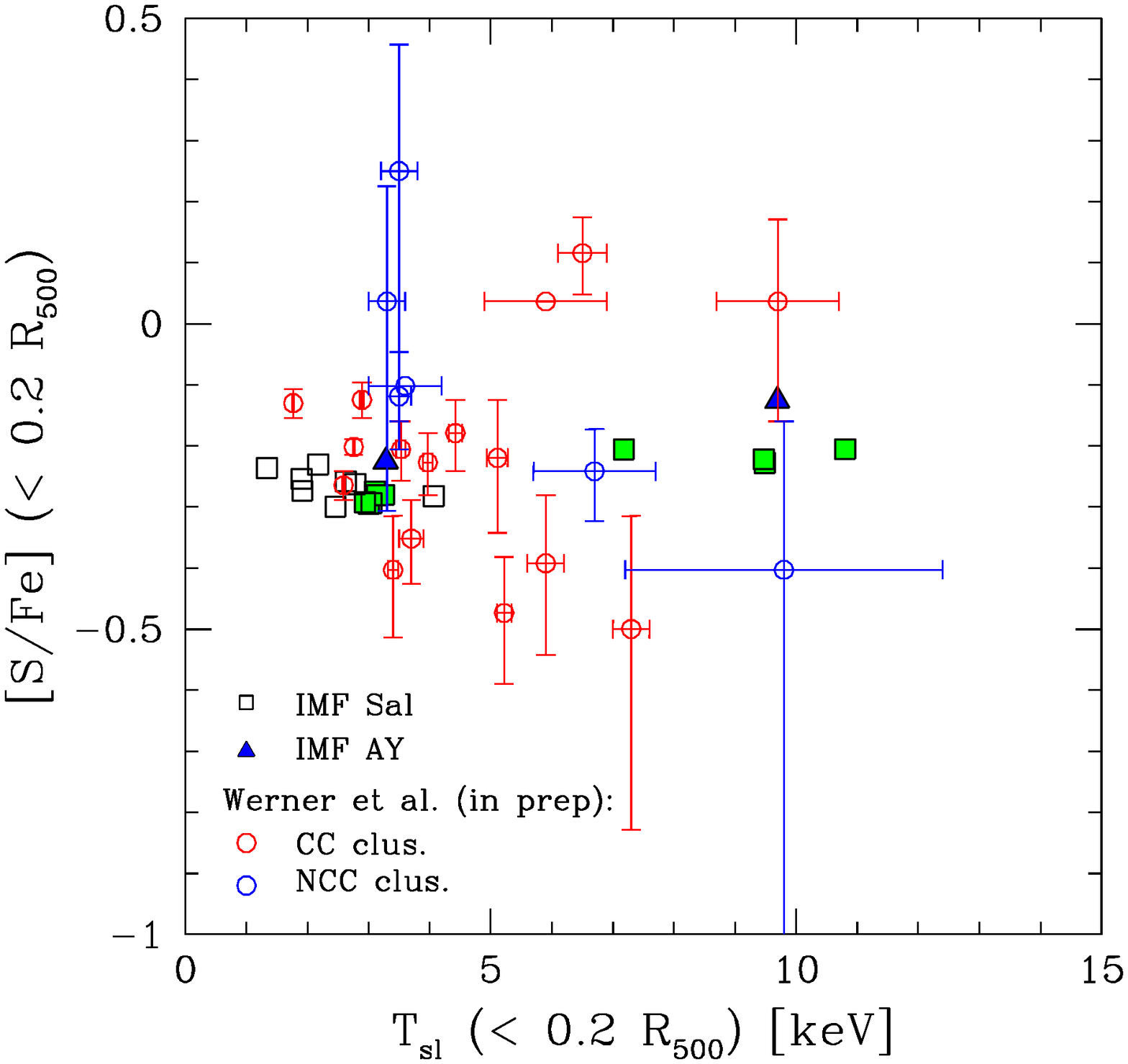}
}
\caption{Left panel: the iron abundance within $0.2r_{500}$ as a
  function of the cluster temperature (here $r_{500}$ is defined as
  the radius encompassing an overdensity of $500\rho_{\rm c}$, with
  $\rho_{\rm c}$ the cosmic critical density). The open squares are for
  simulated clusters analysed by Fabjan et al. (in preparation), which
  assume a Salpeter IMF, while the filled triangles are for two simulated
  clusters which assume the top--heavy AY IMF. Open and filled circles
  with error bars are for cool--core and non cool--core clusters,
  respectively, observed with XMM--Newton by 
  \protect\citet{2007A&A...465..345D} (see also \citealt{werner2008} - Chapter
  16, this volume). Right panel: the same as in the
  left panel, but for the abundance of sulphur relative to iron. All
  abundances have been rescaled to the Solar abundances by  \protect\citet{1998SSRv...85..161G}.}
\label{fi:icmmet}
\end{figure}

Based on a semi--analytical model of galaxy formation,  
\citet{2005MNRAS.358.1247N} carried out a comparison between the
global observed and predicted ICM enrichment for different elements.
In their analysis, these authors considered both the case of a
``standard'' Salpeter IMF and the case in which a top--heavier IMF is
used in starbursts triggered by galaxy mergers. The comparison with
observations, as reported in Fig.~\ref{fi:fig_naga}, led  \citet{2005MNRAS.358.1247N} to conclude that the top--heavier
IMF is required to reproduce the level of enrichment traced by
different elements. All the abundances are predicted to be almost
independent of temperature, at least for $T_X\gtrsim 1$ keV (this
temperature corresponds to the limiting halo circular velocity, above
which gas and metals ejected by superwinds are recaptured by the
halo). While this result agrees with the observational points shown in
the figure, they are at variance with respect to the significant
temperature dependence of the iron abundance found by  \citet{2005ApJ...620..680B}.
 
A different conclusion on the shape of the IMF is instead reached by
other authors. For instance, Fabjan et al. (in preparation) analysed a
set of hydrodynamical simulations of galaxy clusters, which include a
chemical evolution model. In Fig.~\ref{fi:icmmet} we show the result
of this analysis for the comparison between the simulated temperature
dependence of iron (left panel) and sulphur (right panel) and the
results of XMM--Newton observations by 
 \citet{2007A&A...465..345D} (see also \citealt{werner2008}, - Chapter
16, this volume). As for the iron abundance, there is a tendency for
simulations with a Salpeter IMF to overproduce this element, by about
30 per cent. Using a AY top--heavy IMF turns into a much worse
disagreement, with an excess of iron in simulations by a factor
2--3. It is clear that this iron overabundance could be mitigated in
case simulations include some mechanism for diffusion and/or mixing of
metals, which decreases the abundance at small cluster-centric
radii. Of course, this decrease should be such to reproduce the
observed metallicity profiles (see below). As for the [S/Fe] relative
abundance\footnote{Here and in the following, we follow the standard
  bracket notation for the relative abundance of elements $A$ and $B$:
  $[A/B]=\log(Z_A/Z_B)-\log(Z_{A,\odot}/Z_{B,\odot})$.}, the
difference between the two IMFs is much smaller, since the two
elements are produced in similar proportions by different stellar
populations, and, within the relatively large observational
uncertainties, they agree with data.

The comparison between the results of Figs.~\ref{fi:fig_naga} and
\ref{fi:icmmet} is only one example of the different conclusions that
different authors reach on the IMF shape from the study of the
enrichment of the ICM. Tracing back the origin of these differences is
not an easy task. Besides the different descriptions of the relevant
physical processes treated in the SAMs and in the chemo--dynamical
simulations, a proper comparison would also require using exactly the
same model of chemical evolution and the same sets of yields for
different stellar populations. We do recommend that all papers aimed
at modelling the ICM enrichment should carefully include a complete
description of the adopted chemical evolution model.

\section{Metallicity profiles}
\label{profiles}
Observations based on ASCA (e.g.  \citealt{2001ApJ...555..191F}) and Beppo--SAX (\citealt{2004A&A...419....7D}) satellites have revealed for the
first time the presence of significant gradients in the profiles of
ICM metal abundances. The much improved sensitivity of the XMM--Newton
and Chandra satellites are now providing much more detailed
information on the spatial distributions of different metals, thus
opening the possibility of shedding light on the past history of star
formation in cluster galaxies and on the gas-dynamical processes which
determine the metal distribution.

A detailed study of the metal distribution in the ICM clearly requires
resorting to the detailed chemo-dynamical approach offered by
hydrodynamical simulations. In principle, a number of processes,
acting at different scales, are expected to play a role in determining
the spatial distribution of metals: galactic winds (e.g.  \citealt{2004MNRAS.349L..19T,2006MNRAS.371..548R}), transport by buoyant bubbles
 \citep{2007MNRAS.375...15R,2007arXiv0705.2238S}, ram--pressure stripping (e.g.,
 \citealt{2007A&A...466..813K}), diffusion by stochastic gas
motions  \citep{2005MNRAS.359.1041R}, sinking of
highly enriched low--entropy gas \citep{2006MNRAS.368.1540C},
etc. It is clear that  modelling properly all such processes
represents a non--trivial task for numerical simulations, which are
required both to cover a wide dynamical range and to provide a
reliable description of a number of complex astrophysical processes
(\citealt{schindler2008} - Chapter 17, this volume).

\begin{figure}
\centerline{
\hbox{
\includegraphics[height=6cm]{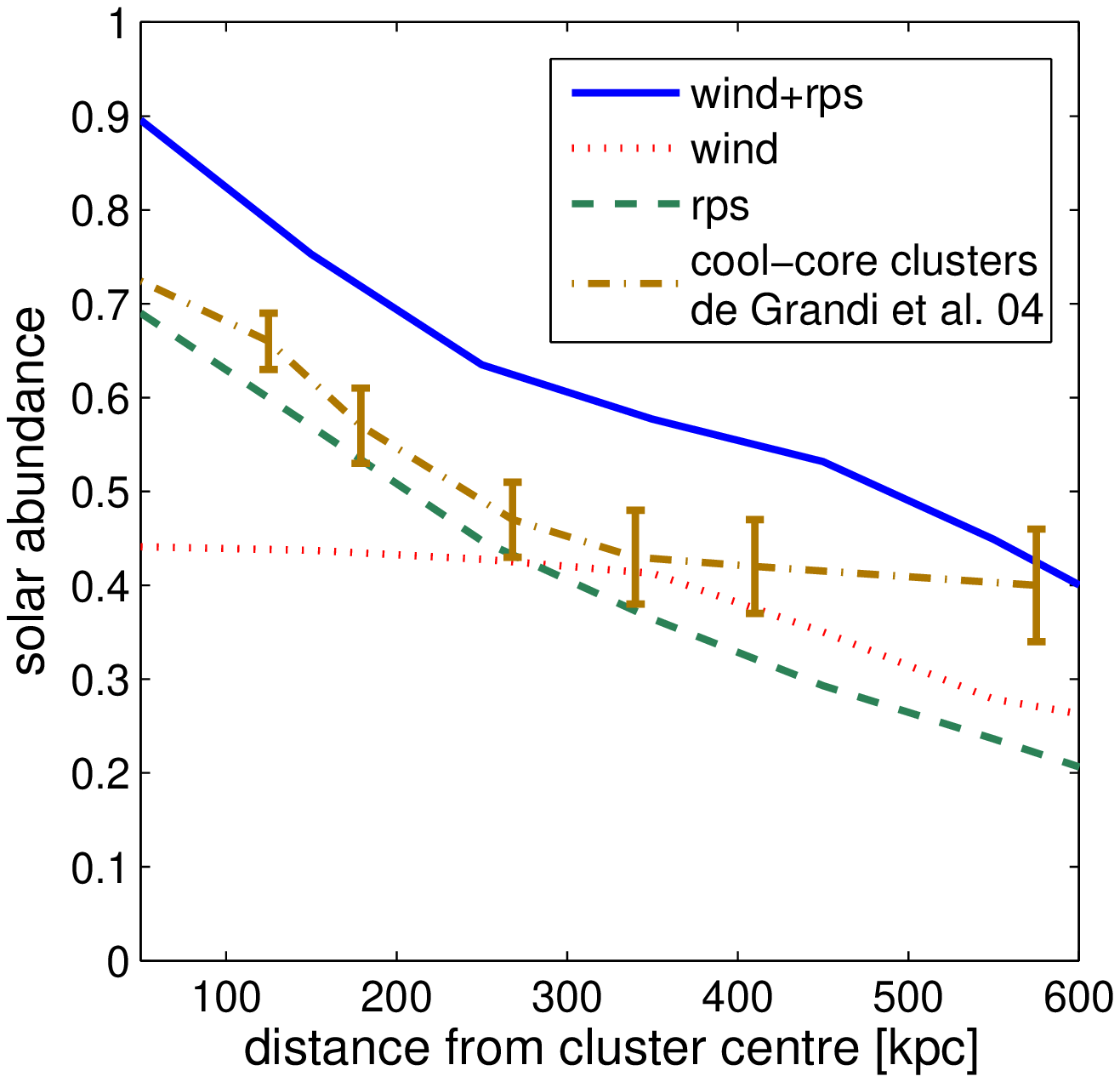}
\includegraphics[height=5.5cm]{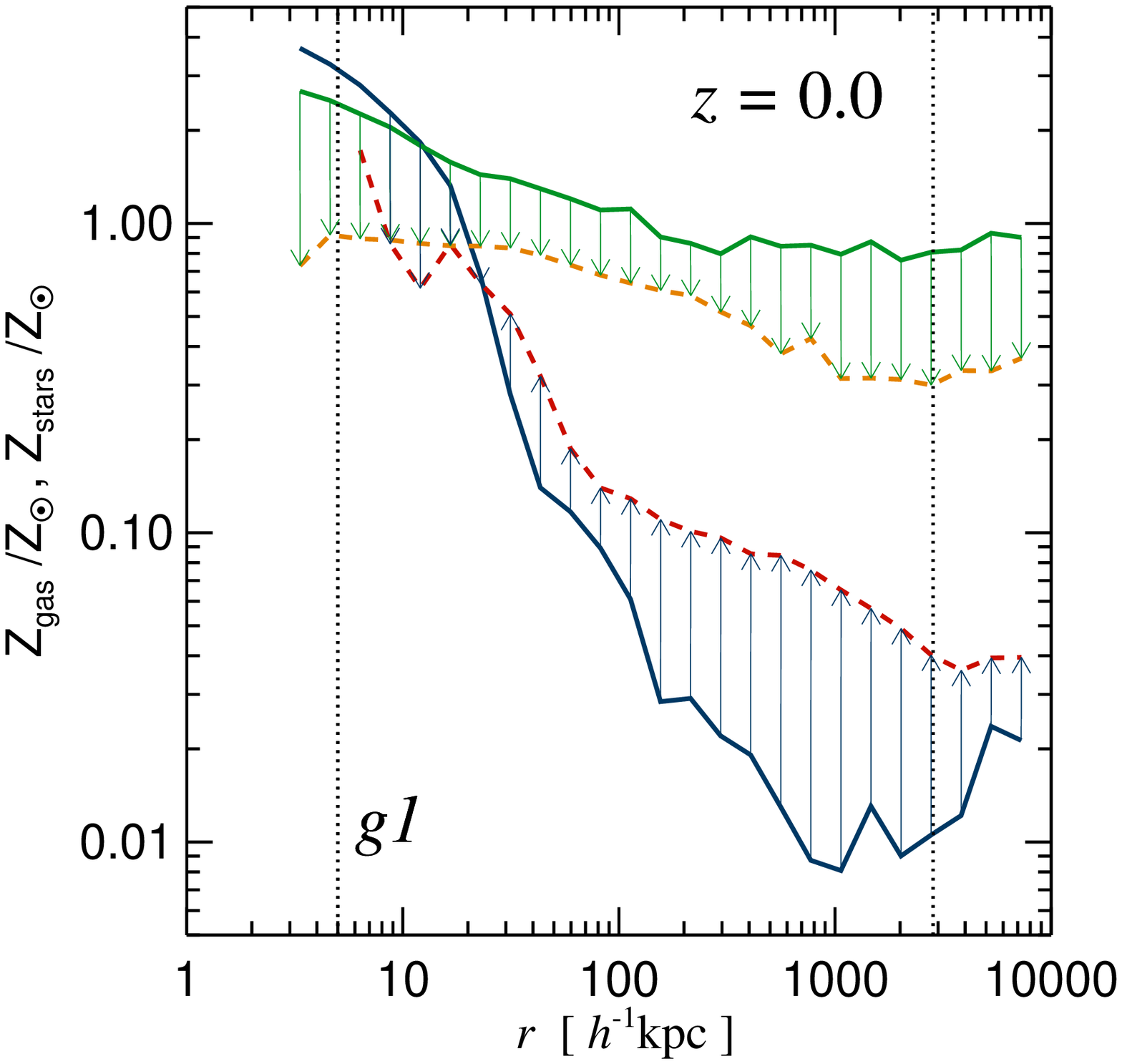}
}}
\caption{Left panel: contribution from ram--pressure stripping and
  from galactic winds to the resulting iron metallicity profile of a
  simulated cluster (from  \protect\citealt{2007A&A...466..813K}). Data points 
  are from the Beppo--SAX observations analysed by  \protect\citet{2004A&A...419....7D}. Right panel: the effect
  of AGN feedback on the profiles of mass-weighted gas metallicity and
  stellar metallicity with simulations of a galaxy cluster (from
 \protect\citealt{2007arXiv0705.2238S}). Continuous lines
  show the metallicity profiles without AGN heating, while dashed
  lines are for the runs with AGN feedback. The arrows illustrate how
  the ICM metallicity is affected by the AGN feedback. The vertical
  dotted lines mark the resolution limit of the simulations and the
  virial radius of the simulated cluster.}
\label{fi:fig_profs1}
\end{figure}

As an example, in the left panel of Fig.~\ref{fi:fig_profs1} we show
the result of the analysis by  \citet{2007A&A...466..813K}, which was aimed at studying the
relative role played by galactic winds and ram--pressure stripping in
determining the metallicity profiles. In these simulations, based on an
Eulerian grid code, these authors use a semi--analytical model to
follow the formation and evolution of galaxies, and described the
evolution of a global metallicity, instead of following different
metal species.
It is interesting to see how different mechanisms contribute in the
simulations to establish the shape of the metallicity profile (see
also \citealt{schindler2008}, - Chapter 17, this volume). In particular,
ram--pressure stripping turns out to be more important in the central
regions, where the pressure of the ICM is higher. On the other hand,
galactic winds are more effective at larger radii, where they can
travel for larger distances, thanks to the lower ICM pressure.

As a further example, we show in the right panel of Fig.~\ref{fi:fig_profs1} the effect of AGN feedback on the distribution of
metals in both gas and stars from the cluster simulations by  \citet{2007arXiv0705.2238S}. These authors included in their
SPH simulations a model for black hole growth and for the resulting
energy feedback. Since these simulations do not include a detailed
model of chemical evolution, the resulting metallicity cannot be
directly compared to observations. Still, this figure nicely
illustrates how the effect of AGN feedback is that of changing the
share of metals between the stars and the gas. Due to the quenching of
low--redshift star formation, a smaller amount of metals are locked
back in stars, with a subsequent reduction of their metallicity.
Furthermore, the more efficient metal transport associated to the AGN
feedback causes a significant increase of the ICM metallicity in the
outer cluster regions, thereby making the abundance profile shallower.

\begin{figure}
\centerline{
\hbox{
\includegraphics[width=0.49\textwidth]{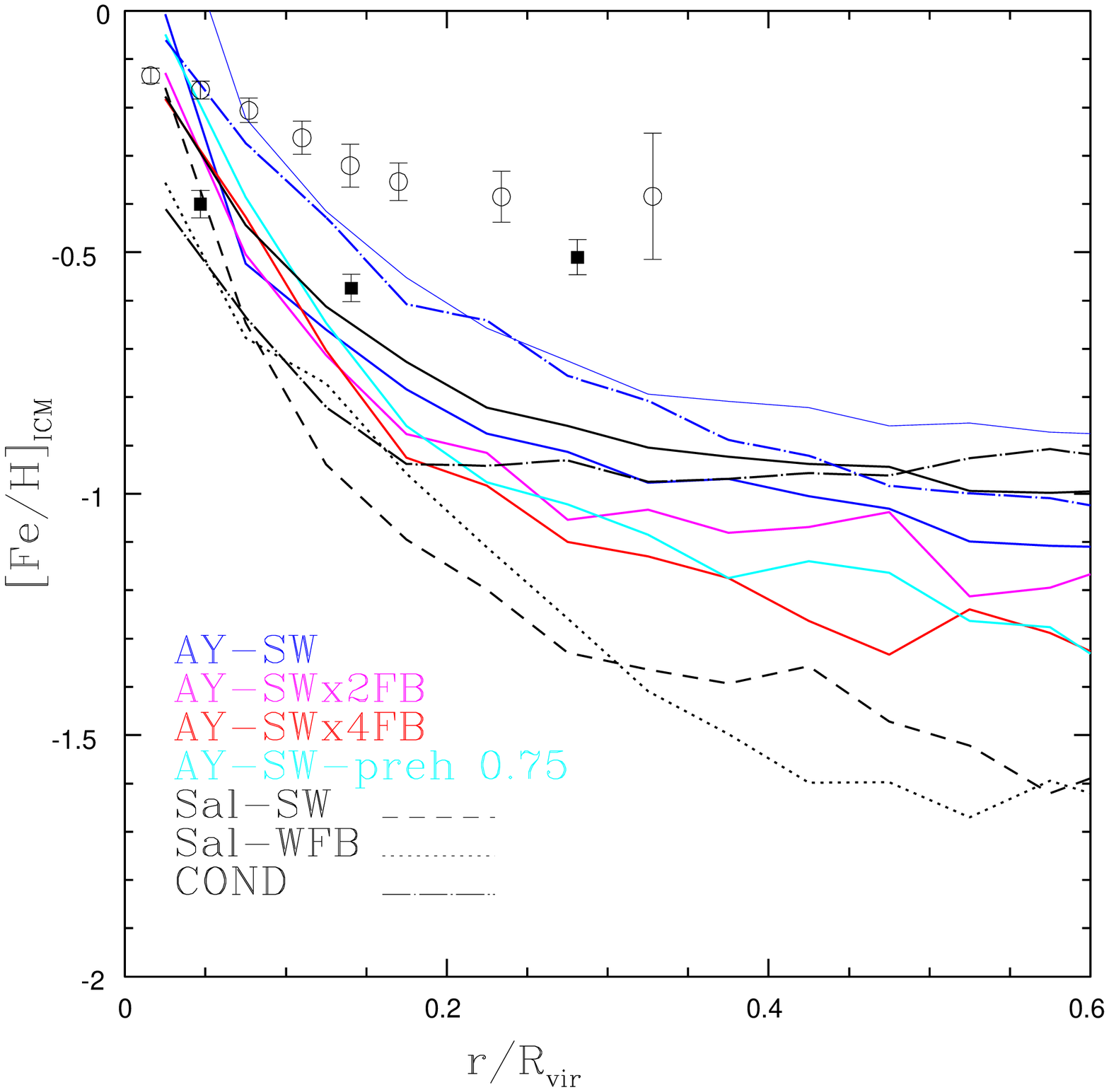}
\includegraphics[width=0.49\textwidth]{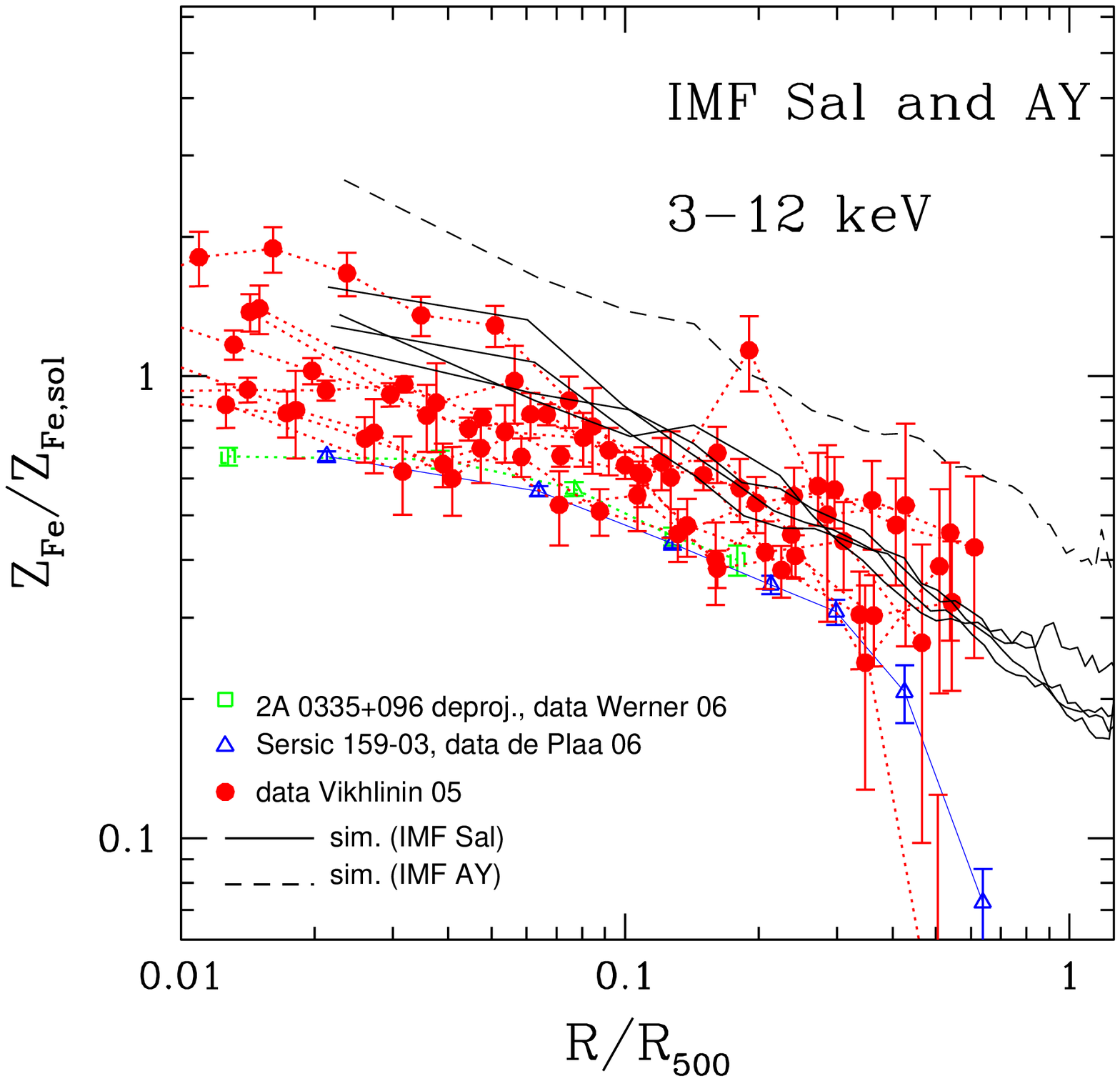}
}}
\caption{Left panel: profiles of iron metallicity at $z = 0$ for
  different runs of two simulated clusters, obtained by changing the
  IMF (AY: Arimoto and Yoshii 1987; Sal: Salpeter 1955), the
  efficiency of the energy feedback and by including the effect of
  thermal conduction (COND; from  \protect\citealt{2005MNRAS.361..983R}). Observational data points
  are taken from  \protect\citet{2004A&A...419....7D}. Right panel: Comparison
  between observed and simulated profiles of iron abundance (from
  Fabjan et al., in preparation). The continuous black curves are for
  four simulated clusters, having virial masses in the range
  $(1-2)\times 10^{15}\msun$, assuming a Salpeter IMF. The black dashed
  line is for one of these clusters, simulated by assuming the
  AY top--heavy IMF. Filled circles with error bars shows observational
  results from Chandra \protect\citep{2005ApJ...628..655V} for clusters
  with temperature above 3 keV, while the open squares and triangles
  are for the profiles of two $\simeq 3$ keV clusters observed with
  XMM--Newton
  \protect\citep{2006A&A...452..397D,2006A&A...449..475W}. The iron
  abundance has been rescaled to the Solar abundances by  \protect\citet{1998SSRv...85..161G}.}
\label{fi:fig_profs2}
\end{figure}

Since iron is the first element for which reliable profiles have been
measured from X--ray observations, most of the comparisons performed
so far with simulations have concentrated on its spatial
distribution. As an example, we show in the left panel of Fig.~\ref{fi:fig_profs2} the comparison between the iron profiles measured
by  \citet{2004A&A...419....7D} and results from
the chemo--dynamical SPH simulations by 
\citet{2006MNRAS.371..548R}. These authors performed simulations for
two clusters having sizes comparable to those of the Virgo and of the
Coma cluster, respectively. After exploring the effect of varying the
feedback efficiency, they concluded that the resulting metallicity
profiles are always steeper than the observed ones. A similar
conclusion was also reached by 
\citet{2003MNRAS.339.1117V}, based on SPH simulations of a larger
ensemble of clusters, which used a rather inefficient form of thermal
stellar feedback. A closer agreement between observed and simulated
profiles of the iron abundance is found by Fabjan et al. (in
preparation, see also  \citealt{Tornatore07}), based on
{\tt GADGET-2} SPH simulations, which include the effect of galactic
winds, powered by SN explosions, according to the scheme presented by
 \citet{2003MNRAS.339..289S}. The results of
this comparison are shown in the right panel of
Fig.~\ref{fi:fig_profs2}. The solid black lines show the results for
simulated clusters of mass in the range $\simeq (1-2)\times
10^{15}\msun$, based on a Salpeter IMF, while the dashed curve is for
one cluster simulated with a top--heavy AY IMF. The observational data
points are for a set of nearby relaxed clusters observed with Chandra
and for two clusters with $T\simeq 3$ keV observed
with XMM--Newton. This comparison shows a
reasonable agreement between simulated and observed profiles, with the
XMM--Newton profiles staying on the low side of the Chandra profiles,
also with a shallower slope in the central regions. While this
conclusion holds when using a Salpeter IMF, a top--heavy IMF is
confirmed to provide too high a level of enrichment.

The different degree of success of the simulations shown in the two
panels of Fig. \ref{fi:fig_profs2} in reproducing observations can
have different origins. The most important reason is probably due to
the different implementations of feedback which, as shown in
Fig. \ref{fi:fig_profs1}, plays a decisive role in transporting metals
away from star-forming regions. Another possible source of difference
may lie in the prescription with which metals are distributed around
star particles. Once again, the details of the numerical
implementations of different effects need to be clearly specified in
order to make it possible to perform a meaningful comparison among
different simulations. Finally, it is also interesting to note that
clusters observed with Chandra seem to have steeper central profiles
than those observed with XMM--Newton and Beppo--SAX. There is no
doubt that increasing the number of clusters with reliably determined
metallicity profiles will help sorting out the origin of such
differences.

\section{Evolution of the ICM metallicity}
\label{evol}
After the pioneering paper based on ASCA data by  \citet{1997ApJ...481L..63M}, the increased sensitivity
of the XMM--Newton and Chandra observatories recently opened the
possibility of tracing the evolution of the global iron content of the
ICM out to the largest redshifts, $z\simeq 1.4$, where clusters have
been secured (e.g.  \citealt{2007A&A...462..429B,2007astro.ph..3156M}). These observations have
shown that the ICM iron metallicity decreases by about 50 per cent
from the nearby Universe to $z\simeq 1$. This increase of metallicity
at relatively recent cosmic epochs can be interpreted as being due
either to a fresh production of metals or to a suitable
redistribution of earlier-produced metals. Since SN\,Ia, which provide
a large contribution to the total iron budget, can release metals over
fairly long time scales, the increase of the iron abundance at $z<1$
is not necessarily in contradiction with the lack of recent star
formation in the bulk of cluster galaxies. On the one hand, recent
analyses \citep{2005MNRAS.362..110E,2006ApJ...648..230L} have shown that the observed metallicity
evolution is consistent with the observed SN\,Ia rate in clusters
\citep{2002MNRAS.332...37G}. On the other hand, a number of processes,
such as stripping of metal enriched gas from merging galaxies (e.g.,
 \citealt{2007MNRAS.tmpL..43C}) or sinking of highly
enriched low--entropy gas towards the cluster central regions 
\citep{2006MNRAS.368.1540C} may also significantly contribute to the
ICM metallicity evolution, through a redistribution of enriched gas.
Indeed, the amount of accreted gas since $z=1$ is at least comparable
to that present within the cluster at that redshift. Therefore, the
evolution of the ICM metallicity is expected to be quite sensitive to
the enrichment level of the recently accreted gas.  On the other hand,
as noted by  \citet{1997ApJ...488...35R}, if ram--pressure
stripping plays a dominant role in determining the ICM enrichment, one
should then expect to observe more massive clusters, whose ICM has a
higher pressure, to be more metal rich than poorer systems. In fact,
observations do not show this trend and, if any, suggest a decrease of
the iron abundance with the ICM temperature (e.g.,  \citealt{2005ApJ...620..680B}).

\begin{figure}
\includegraphics[width=\textwidth]{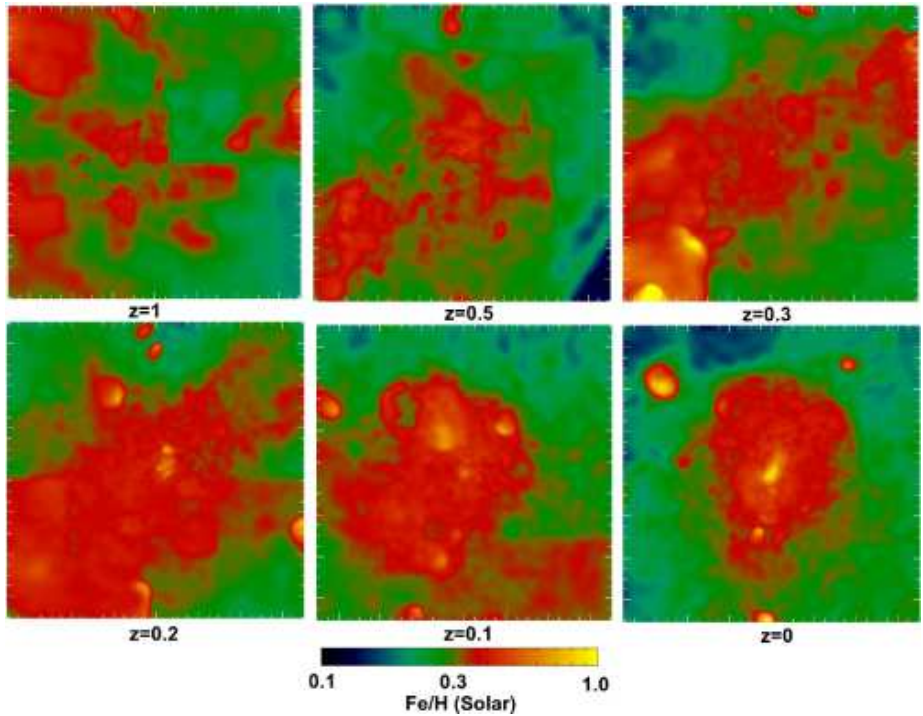}
\caption{Evolution of the iron abundance within 700 $h^{-1}$~kpc from
  the centre of a simulated cluster at $z=1,0.5,0.3,0.2,0.1,0$, from
  upper left to bottom right panel. The maps show the projection of
  the mass--weighted iron abundance by number relative to hydrogen,
  Fe/H, with respect to the solar value (adapted from 
  \protect\citealt{2006MNRAS.368.1540C}).}
\label{fi:fig_cora}
\end{figure}

\citet{2006MNRAS.368.1540C} used a hybrid approach based on coupling a
SAM model of galaxy formation to a non--radiative hydrodynamical
simulation to study the gas dynamical effects which lead to
the re-distribution of metals from the star forming regions. The
results of her analysis are summarised in the metallicity maps shown
in Fig.~\ref{fi:fig_cora}. Metals are produced inside small halos
which at high redshift define the proto--cluster region. Since the gas
in these halos has undergone no significant shocks by diffuse
accretion, its entropy is generally quite low. This low--entropy
highly enriched gas is therefore capable to sink towards the central
regions of the forming cluster. As a consequence, this leads to an
increase of the metallicity in the regions which mostly contribute to
the emissivity, thereby causing a positive evolution of the observed
metallicity.
 
\begin{figure}
\centerline{
\hbox{
\includegraphics[width=0.49\textwidth,height=6cm]{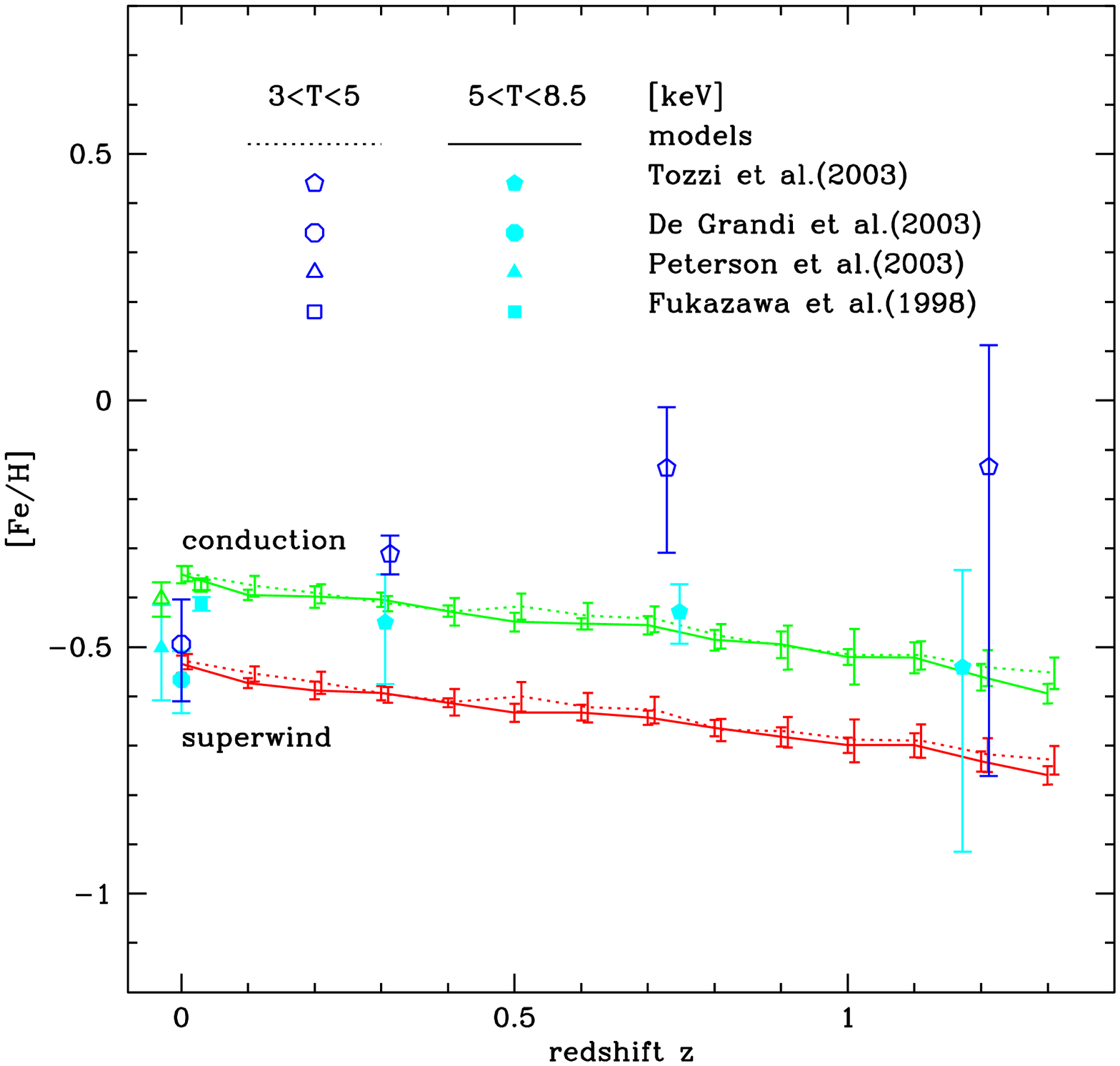}
\includegraphics[width=0.49\textwidth,height=6.5cm]{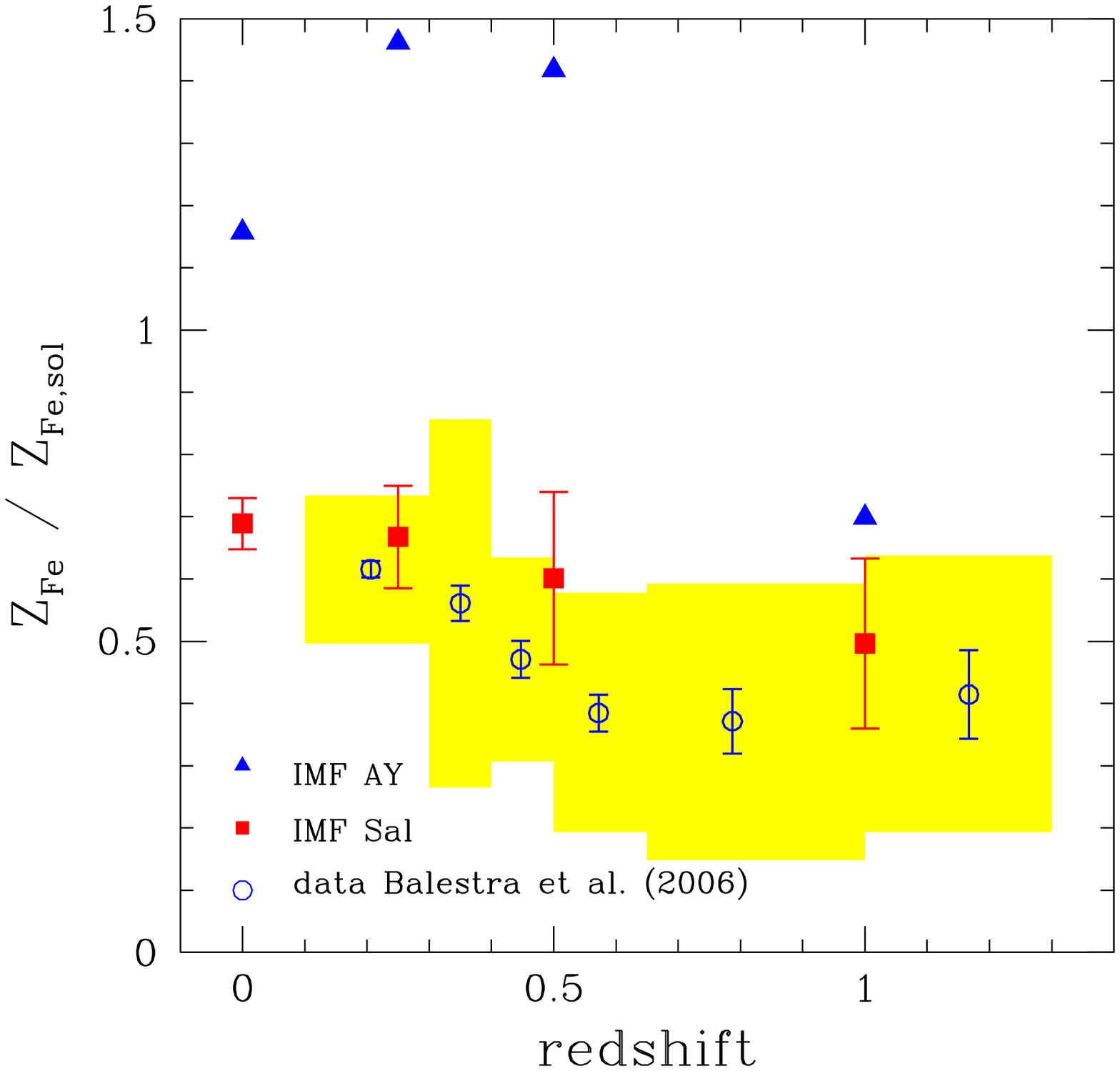}
}}
\caption{The redshift evolution of the iron abundance. Left panel:
  predictions by different flavours of the SAM by Nagashima et
  al. (2005), based on a top--heavy IMF, compared with observational
  data from different sources, for clusters within two temperature
  ranges. Error bars on model predictions are the 1$\sigma$ scatter
  over the ensemble of cluster formation histories. Right panel:
  comparison between observational data and full hydrodynamical
  simulations by Fabjan et al. (2007, in preparation). Open circles
  come from the analysis of Chandra data by  \protect\citet{2007A&A...462..429B}. The shaded area
  represents the r.m.s. scatter within each redshift interval, while
  the error bars are the 1$\sigma$ confidence level. The filled squares
  are for a set of four simulated clusters,
  using a Salpeter IMF, with temperature ${\mathrm k}T> 5$ keV, with error bars
  indicating the corresponding 1$\sigma$ scatter. The filled triangles
  refer to only one cluster, out of four,
  simulated with the top--heavy IMF. In both panels, the values of the
  iron abundance have been rescaled to the Solar abundances by
  \protect\citet{1998SSRv...85..161G}.}
\label{fi:evol}
\end{figure}

\citet{2005MNRAS.358.1247N} used a SAM approach,
not coupled to hydrodynamical simulations, to predict the evolution of
the ICM metallicity. In this case, the evolution is essentially driven
by the low--redshift star formation, since any redistribution of
enriched gas is not followed by the model. The result of their
analysis (left panel of Fig.~\ref{fi:evol}) shows that a very mild
evolution is predicted. A prediction of the evolution of the iron
abundance from hydrodynamical simulations, which self--consistently
follow gas cooling, star formation and chemical evolution, has been
more recently performed by Fabjan et al. (in preparation). The results
of their analysis are shown in the right panel of Fig.~\ref{fi:evol},
where they are also compared to the observational results by  \citet{2007A&A...462..429B}. Also in this case, the results
based on a Salpeter IMF are in reasonable agreement with observations,
while a top--heavier IMF produces too high metallicities at all
redshifts. Quite interestingly, at low redshift the run with
top--heavy IMF shows an inversion of the evolution. This is due to the
exceedingly high star formation triggered by heavily enriched gas,
which locks back in stars a substantial fraction of the gas metal
content in the cluster central regions. This highlights the crucial
role played by star formation in both enriching the ICM and in locking
back metals in the stellar phase. This further illustrates how the
resulting metallicity evolution must be seen as the result of a
delicate interplay of a number of processes, which all need to be
properly described in a self--consistent chemo--dynamical model of the
ICM.

\section{Properties of the galaxy population}
\label{galaxies}
The physical processes which determine the thermodynamical and
chemical properties of the ICM are inextricably linked to those
determining the pattern of star formation in cluster galaxies. For
this reason, a successful description of galaxy clusters requires a
multiwavelength approach, which is able at the same time to account
for both the X--ray properties of the diffuse hot gas and of the
optical/near--IR properties of the galaxy populations. In this sense,
the reliability of a numerical model of the ICM chemical enrichment is
strictly linked to its ability to reproduce the basic properties of
the galaxy population, such as luminosity function and
colour--magnitude diagram. While the study of the global properties of
the galaxy populations has been conducted for several years with
semi--analytical models of galaxy formation  \citep{2004MNRAS.349.1101D,2005MNRAS.358.1247N}, it is only quite recently that
similar studies have been performed with cosmological hydrodynamical
simulations, thanks to the recent advances in supercomputing
capabilities and code efficiency.

\begin{figure}
\centerline{
\hbox{
\includegraphics[width=0.37\textwidth]{fig11a.ps}
\includegraphics[width=0.61\textwidth]{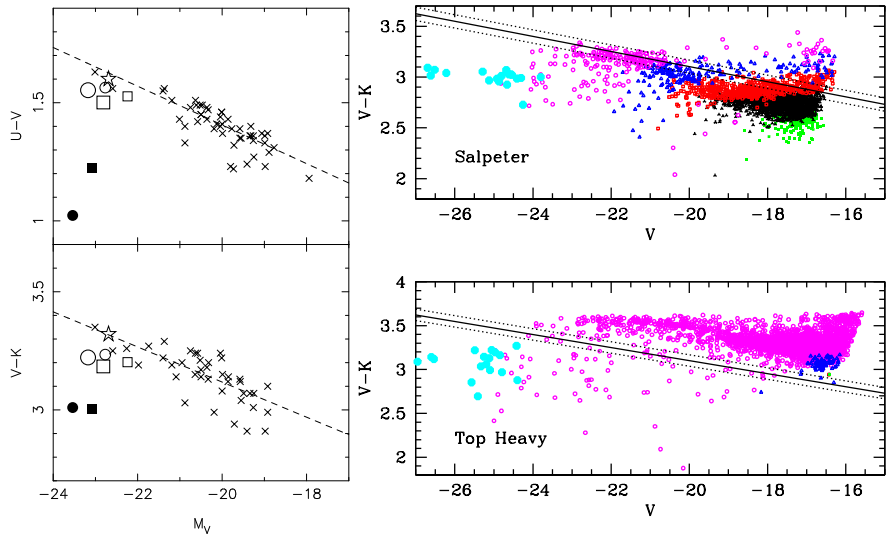}
}}
\caption{Left panel: Colour--magnitude relations for the simulations of
  elliptical galaxies \protect\citep{2003MNRAS.346..135K}, based on
  a Salpeter IMF, compared to the observational results for the
  galaxies in Coma \protect\citep{1992MNRAS.254..601B} (crosses) and
  for NGC 4472 (open star). The filled square and circle are the
  results for the reference runs, whereas the open symbols are when the
  contribution from stars younger than 2 or 8 Gyr are ignored. Right
  panel: the colour--magnitude relation for simulated clusters
  (symbols). Upper and lower panels refer to simulations performed
  with a Salpeter IMF and with the AY IMF, respectively (from  \protect\citealt{2006MNRAS.373..397S}). The big dots are for the
  BCGs of the simulated clusters. The straight lines indicate the
  best-fitting relation and the scatter observed for the Coma
  galaxies.}
\label{fi:gals}
\end{figure}

As we have already mentioned, hydrodynamical simulations treat the
process of star formation through the conversion of cold and dense gas
particles into collisionless star particles. Each star particle is
then characterised by a formation epoch and a metallicity. Therefore,
it can be treated as a SSP, for which luminosities in different bands
can be computed by resorting to suitable stellar population synthesis
models. Since the colours of a SSP are quite sensitive to metallicity,
it is necessary for simulations to include a detailed chemical
evolution model in order to reliably compare predictions on
optical/near--IR properties of galaxies with observational data. This
approach has been pursued by 
\citet{2003MNRAS.346..135K}, who performed simulations of elliptical
galaxies and by  \citet{2005MNRAS.361..983R} and \citet{2006MNRAS.373..397S}, who performed simulations of galaxy
clusters. The results of these analyses are summarised in Fig.~\ref{fi:gals}. In the left panels we show the position on the
colour--magnitude diagram of the elliptical galaxies simulated by
 \citet{2003MNRAS.346..135K}. This plot clearly
indicates that elliptical galaxies are generally much bluer than
observed, as a result of an excess of recent star formation, which is
not properly balanced by an efficient feedback mechanism. Indeed, it
is only after excluding the contribution from recently formed stars
that \citet{2003MNRAS.346..135K} were able to
obtain ellipticals with more realistic colours.

In the right panels we show the colour--magnitude relation (CMR) of the
galaxy population of an extended set of clusters simulated by  \citet{2006MNRAS.373..397S} based both on a Salpeter IMF and on
a top--heavy IMF. Quite apparently, the bulk of the galaxy population
traces a CMR which is consistent with the observed one, at least for a
Salpeter IMF.  As for the top--heavy IMF, it predicts too red colours
as a consequence of a too high metal content of galaxies. Also
interesting to note, brighter redder galaxies are on average more
metal rich, therefore demonstrating that the CMR corresponds in fact
to a metallicity sequence (see also  \citealt{2005MNRAS.361..983R}). However, a clear failure of these
simulations is that the BCGs are always much bluer than observed. This
result agrees with the exceedingly blue colour of simulated
ellipticals shown in the left panels. This is due to
an excess of recent star formation, which, in BCGs of $\sim
10^{15}M_\odot$ simulated clusters, can be as high as
500--1000~M$_\odot$\,yr$^{-1}$. 
As discussed by \citealt{borgani2008} - Chapter 13, this
volume, the presence of overcooling in cluster simulations manifests
itself with the presence of steep negative temperature gradients in the
central regions.  In this sense, wrong BCG colours and temperature
profiles in core regions are two aspects of the same problem, i.e. the
difficulty of balancing the cooling runaway within simulated clusters.

Since all these simulations include recipes for stellar feedback,
although with different implementations and efficiencies, the solution
to the overcooling in clusters should be found in some non--stellar
feedback, such as that provided by AGN. Recent studies of the galaxy
properties, based on semi--analytical models, have shown that an
energy feedback of the sort of that expected from AGN, is able to
suppress the bright end of the galaxy luminosity function, which is
populated by the big ellipticals residing at the centre of clusters
(e.g.,  \citealt{2006MNRAS.370..645B,2006MNRAS.365...11C}). There is no doubt that providing
robust predictions on the properties of the galaxy population will be
one of the most important challenges for simulations of the next
generation to be able of convincingly describing the thermo-
and chemo-dynamical properties of the hot gas.

\section{Summary}
\label{summary}
The study of the enrichment pattern of the ICM and its evolution
provides a unique means to trace the past history of star formation and
the gas--dynamical processes, which determine the evolution of the
cosmic baryons. In this overview, we discussed the concept of model of
chemical evolution, which represents the pillar of any
chemo--dynamical model, and presented the results obtained so far in
the literature, based on different approaches.  Restricting the
discussion to those methods which follow the chemical enrichment
during the cosmological assembly of galaxies and clusters of galaxies,
they can be summarised in the following categories.\\
{\bf (1)} Semi--analytical models (SAMs) of galaxy formation
\citep{2004MNRAS.349.1101D,2005MNRAS.358.1247N}: the production of
metals is traced by the history of galaxy formation within the
evolving DM halos and no explicit gas--dynamical description of the
ICM is included. These approaches are computationally very cheap and
provide a flexible tool to explore the space of parameters determining
galaxy formation and chemical evolution. A limitation of these
approaches is that, while they provide predictions on the global metal
content of the ICM, the absence of any explicit gas dynamical
description causes the lack of useful information
on the spatial distribution of metals.\\
{\bf (2)} SAMs coupled to hydrodynamical simulations. In this case,
the galaxy formation is followed as in the previous approach, but the
coupling with a hydrodynamical simulation allows one to trace the fate
of the enriched gas and, therefore, to study the spatial distribution of
metals. Different authors have applied different implementations of
this hybrid approach, by focusing either on the role of the chemical
evolution model \citep{2006MNRAS.368.1540C} or on the effect of
specific gas--dynamical processes, such as ram--pressure stripping and
galactic winds, in enriching the ICM (e.g.,  \citealt{2007A&A...466..813K}). The advantage of this approach is
clearly that gas--dynamical processes are now included at some
level. However, the galaxy formation process is not followed in a
self--consistent way from the cooling of the gas during the cosmic
evolution.\\
{\bf (3)} Full hydrodynamical simulations, which self--consistently
follow gas cooling, star formation and chemical evolution
\citep{2003MNRAS.339.1117V,2003MNRAS.346..135K,2005MNRAS.361..983R,Tornatore07}. Rather
than being described by an external recipe, galaxy formation is now
the result of the cooling and of the conversion of cold dense gas into
stars, as implemented in the simulation code. The major advantage of
this approach is that the enrichment process now comes as the result
of the star formation predicted by the hydrodynamical simulation,
without resorting to any external model.  The bottleneck, in this
case, is represented by the computational cost, which becomes
prohibitively high if one wants to resolve the whole dynamic range
covered by the processes of metal production and distribution,
feedback energy release, etc. For this reason, these simulation
codes also need to resort to sub--resolution models, which provide an
effective description of a number of physical processes. However,
there is no doubt that the ever improving supercomputing technology
and code efficiency make, in perspective, this approach the way to
study the past history of cosmic baryons.

As already emphasised, developing a reliable model of the cosmic history
of metal production is a non--trivial task. This is due to the
sensitive dependence of the model predictions on both the parameters
defining the chemical evolution model (i.e., IMF, stellar life--times,
stellar yields, etc.), and on the processes, such as ejecta from SN
explosions and AGN, stripping, etc., which at the same time regulate
star formation and transport metals outside galaxies. 
In the light of these difficulties, it is quite reassuring that
the results of the most recent hydrodynamical simulations are in
reasonable agreement with the most recent observational data from
Chandra and XMM--Newton. 

In spite of this success, cluster simulations of the present
generation still suffer from important shortcomings. The most important
of them is probably represented by the difficulty in regulating
overcooling at the cluster centre, so as to reproduce at the same time
the observed ``cool cores'' and the passively evolving massive
ellipticals. Besides improving the description of the relevant
astrophysical processes in simulation codes, another important issue
concerns understanding the possible observational biases which
complicate any direct comparison between data and model predictions
\citep{2007arXiv0707.1573K,2007arXiv0707.2614R}. In this respect,
simulations provide a potentially ideal tool to understand these
biases. Mock X--ray observations of simulated clusters, which include
the effect of instrumental response, can be analysed exactly in the
same way as real observational data. The resulting ``observed''
metallicity can then be compared with the true one to calibrate out
possible systematics. There is no doubt that observations and
simulations of the chemical enrichment of the ICM should go
hand in hand in order to fully exploit the wealth of information
provided by X--ray telescopes of the present and of the next
generation.

\begin{acknowledgements}
The authors thank ISSI (Bern) for support of the
team ``Non--virialized X-ray components in clusters of galaxies''.  We
would like to thank Cristina Chiappini, Francesca Matteucci, Simone
Recchi and Paolo Tozzi for a number of enlightening discussions. We
also thank Florence Durret for a careful reading of the manuscript and
Norbert Werner for providing the observational data points in Fig.~5
prior to publication. The authors acknowledge support by the PRIN2006
grant ``Costituenti fondamentali dell'Universo'' of the Italian
Ministry of University and Scientific Research, by the Italian
National Institute for Nuclear Physics through the PD51 grant, by the
Austrian Science Foundation (FWF) through grants P18523-N16 and
P19300-N16, and by the Tiroler Wissenschaftsfonds and through the
UniInfrastrukturprogramm 2005/06 by the BMWF.
\end{acknowledgements}

\bibliographystyle{aa}
\bibliography{18_borgani}

\end{document}